\documentclass[fleqn,usenatbib]{mnras}

\usepackage{newtxtext,newtxmath}
\usepackage[T1]{fontenc}
\usepackage{ae,aecompl}

\usepackage{graphicx, subfigure}	% Including figure files
\usepackage{amsmath}	% Advanced maths commands
\usepackage{amssymb}	% Extra maths symbols
\usepackage{bm}
\usepackage{comment}
\usepackage{multirow}
\usepackage[utf8]{inputenc}
\title[High-resolution survey for planets in Taurus]{High-resolution survey for planetary companions to young stars in the Taurus Molecular Cloud}

\author[A.~L. Wallace et al.]{A.~L. Wallace$^{1}$\thanks{E-mail: alexander.wallace@anu.edu.au}, J.~Kammerer$^{1,2}$, M.~J.~Ireland$^{1}$, C.~Federrath$^{1}$, \newauthor A.~L.~Kraus$^{3}$, S.~T.~Maddison$^{4}$, A.~C.~Rizzuto$^{3}$, E.~K.~Birchall$^{1}$, F.~Martinache$^{5}$
\\
$^{1}$Research School of Astronomy and Astrophysics, Australian National University, Canberra, ACT 2611, Australia\\
$^{2}$European Southern Observatory, Karl-Schwarzschild-Str 2, 85748, Garching, Germany\\
$^{3}$Department of Astronomy, University of Texas, Austin, TX 78712, United States of America\\
$^{4}$Centre for Astrophysics and Supercomputing, Swinburne University of Technology, Melbourne, VIC 3122, Australia\\
$^{5}$Laboratoire Lagrange, Universit\'e C\^ote d'Azur, Observatoire de la C\^ote d'Azur, CNRS, Parc Valrose, B\^at. H. FIZEAU, 06108 Nice, France}

\date{Accepted XXX. Received YYY; in original form ZZZ}

\pubyear{2015}

\begin{document}
\label{firstpage}
\pagerange{\pageref{firstpage}--\pageref{lastpage}}
\maketitle

\begin{abstract}
Direct imaging in the infrared at the diffraction limit of large telescopes is a unique probe of the properties of young planetary systems.
We survey 55 single class I and class II stars in Taurus in the L’ filter using natural and laser guide star adaptive optics and the near-infrared camera (NIRC2) of the Keck~II telescope, in order to search for planetary-mass companions.
We use both reference star differential imaging and kernel phase techniques, achieving typical 5-sigma contrasts of $\sim$6 magnitudes at separations of 0.2'' and $\sim$8 magnitudes beyond 0.5''.
Although we do not detect any new faint companions, we constrain the frequency of wide separation massive planets, such as HR 8799 analogues. We find that, assuming hot-start models and a planet distribution with power-law mass and semi-major axis indices of -0.5 and -1, respectively, less than 20\% of our target stars host planets with masses >2\,M$_{\rm{J}}$ at separations >10\,au.
%Young equivalents of HR8799bc would have been detectable in approximately half of our systems, and HR8799de in $\sim$20\% of our systems.
%However, recent observationally constrained core accretion models produce giant planets that are unresolved by 8-10~m class telescopes even in the nearest star-forming regions. 
%We argue that a significant increase in angular resolution, e.g., using the VLTI, is needed to find the expected population of luminous young giant planets.
\end{abstract}


%%%%%%%%%%%%%%%%%%%%%%%%%%%%%%%%%%%%%%%%%%%%%%%%%%

%%%%%%%%%%%%%%%%% BODY OF PAPER %%%%%%%%%%%%%%%%%%

\section{Introduction}
Direct imaging of exoplanets is an important method to study planetary systems and gain insight into formation scenarios.  Most directly imaged exoplanets have been found in young star systems when the planets are still hot and emit in the infrared (e.g.  HR 8799 \citep{marois2008direct}) while some have been found in the process of formation \citep{keppler2018discovery}.  Most directly imaged planets are at wide separations ($>20$\,au) from their host stars but models of planet distributions \citep{fernandes2019hints} indicate that these systems are rare.  Giant planets such as Jupiter are likely to form by core accretion which occurs closer to the star ($\sim5$\,au.)

The Taurus Molecular Cloud (TMC) is ideal for studying planet formation due to its relative proximity ($\sim140\,$pc) and numerous young stars ($<2\,$Myr) \citep{torres2009vlba}. Many of these young stars have prominent disc structures \citep{partnership2015first, huang2020multifrequency}, which may be indicative of planet formation. A planet in the process of formation will radiate in the near-infrared. In an optically thick disc, the planet will be hidden at these wavelengths.  However, a giant planet ($\sim0.5\,$M$_{\rm{J}}$ and above) is expected to clear a gap in the disc \citep{crida2007cavity}.  Many of the discs in our sample have gaps present in their dust distribution, as indicated by ALMA surveys \citep{long2018gaps} and, although their origin is still hotly debated, one possibility is giant planet formation.

The circumstellar discs in the TMC have been extensively studied over the years in terms of their mass \citep{andrews2005circumstellar,andrews2013mass}, structure and distribution, as have the discs in other nearby star-forming regions such as Upper Scorpius and Ophiuchus \citep{carpenter2014alma,van2016dust,kuruwita2018multiplicity}.  Surveys have also been conducted to detect planets in these star-forming regions \citep{tanner2007sim,metchev2009palomar} and some have found potentially planet-mass companions at wide separations (e.g., DH~Tau~b; \citet{itoh2005young}) as well as a close companion to CI Tau using radial velocity \citep{johns2016candidate}.  However, these surveys were unable to achieve the necessary sensitivity for planetary-mass companions on solar-system scales.  \citet{kraus2011mapping} managed to detect new brown dwarf companions at small separations and achieved a mass sensitivity of $\sim20\,$M$_{\rm{J}}$. In part of this earlier work, emission with total luminosity comparable to a forming planet was discovered around LkCa~15 \cite{kraus2011lkca}, although the complex transitional (or ``pre-transitional'') nature of this disc has meant that a physically motivated radiative transfer model could not be made at the time. A scattering origin for the emission was, however, strongly suggested by further observations with Sphere and ZIMPOL \citep{Thalmann15}.

The purpose of our study is to search for giant planets around young stars in the TMC still accreting from their discs, and determine dominant mechanisms for planet formation.  This study is unique because, for the first time, we attempted to probe the inner regions of these systems in search of solar-system analogues, at a time where the significant presence of disc gas means that planetary luminosities would be highest.  As the TMC is the nearest star-forming region of its size \citep{gudel2007xmm} and the projected separation of any planets decreases with the distance, it is the most favourable region for resolving the peak of the giant planet distribution at physical separations of $<10$\,au.  Using planet distributions from \citet{cumming2008keck} and later by \citet{fernandes2019hints}, it is clear that even at this close distance, the majority of planets are inside the $\sim$20-30\,au limits of a typical coronagraph.

When planets form, they heat up and radiate in infrared wavelengths and are at their brightest during runaway accretion.  After formation, the planets continue to radiate for some time and should still be self luminous after millions of years as shown by the HR~8799 and $\beta$-Pictoris systems \citep{marois2008direct,lagrange2010giant}.  The evolution of planet luminosity is an important factor in this work as it determines our detection capability. However, many details of the accretion luminosity remain uncertain. The luminosity of a circum-planetary disc is dependant on the accretion rate as well as the mass and radius of the planet, and whether and where the circum-planetary disc is truncated \citep{zhu2015accreting}. The post-accretion luminosity of hot-start planets (i.e., planets which do not lose entropy in an accretion shock) has been modelled for some time as applied to brown dwarfs \citep[e.g.][]{baraffe2003evolutionary}. ``Cold'' start models, where all accretion shock luminosity is radiated away, can have very different initial luminosities, especially for massive planets \cite[e.g.][]{marley2007luminosity}, although detailed shock models considering radiative transfer and reasonable accretion rates have recently shown that ``warm'' start models are much more realistic \citep{Marleau19}. Additionally, models of post-shock gas has shown a zone of stability with intial entropies around 10--11\,$k_\mathrm{B}$/baryon, termed ``stalling'' accretion \citep{berardo2017evolution}. 
 
Planets cool and fade as they age but the cooling time is very dependent on the mass and internal entropy of the planet, with high-entropy low-mass planets cooling the fastest, and e.g. a 5\,M$_J$ planet cooling at 0.5\,$k_\mathrm{B}$/baryon/Myr from an initial internal entropy of 1.5\,$k_\mathrm{B}$/baryon \citep{spiegel2012spectral}. Irrespective of these uncertainties in post-formation luminosity evolution, the best time to directly image exoplanets is shortly after their formation, when they are at their highest luminosity. The canonically young age of the TMC provides a perfect environment in which to search for these planets.

In Section~\ref{sec:sample} we describe our survey sample of 55 stars in the TMC using the Near Infrared Camera (NIRC2) on the Keck~II telescope in 2015 and 2016.  In Section~\ref{sec:observations_analysis} we describe our observation, data reduction and small angle analysis methods.  In Section~\ref{sec:wide} we expand our analysis to wider separations and identify companions.  Section~\ref{sec:freq} combines our methods for all separations to place limits on the frequency of wide planets.   Our conclusions are presented in Section~\ref{sec:conclusion}.

% ================================
\section{Survey sample}
\label{sec:sample}
In choosing our sample of stars in the TMC, we decided to only select single stars, which we define as stars with no known stellar companion within 1''.  The reason for this is firstly, that multiple star systems can cause issues with adaptive optics but also for reasons of simplicity.  The data reduction and analysis is simplified if there is only one bright central star to consider and there is less theoretical complexity regarding models of planet formation.  An exception to this is V410 Tau \citep{ghez1997high}, which we use to ensure we are correctly oriented and to verify our data processing pipelines.  Note that we include close ($\la$1\,au) spectroscopic binaries in our sample if they meet all other criteria, as we argue wide companions in these systems are likely to be unaffected by the dynamics of the close orbit. Two known systems are in our sample: DQ Tau \citep{Mathieu97} and UZ Tau A \citep{Prato02}.

We primarily consider class II targets because, at this stage in stellar evolution, circum-stellar discs have been observed to have very low mass, between 0.2\% and 0.6\% of the host-star mass \citep{andrews2013mass}. This indicates, if there is planet formation, the most massive planets will have already formed by this phase.  We also consider class I objects such as HL Tau, which has a circumstellar disc containing notable gaps and rings, which may be indicative of planet formation \citep{brogan20152014}.  Our targets were taken from \citet{kraus2011mapping}.  We selected targets based on their J-K magnitude colours and only selected targets with J-K$<$4 and K magnitude $<$10 which can be used as a guide for the approximate L' magnitude.  We also made a cut on the spectral type, excluding targets listed as later than M3 in \citet{kraus2011mapping}. This cutoff was chosen to include the relatively abundant low-mass stars in the TMC while cutting-out stars that would be too faint for AO observations and too low in mass to expect giant planets. We note that recent studies have produced updated spectral types. The spectral types shown in Table~\ref{tab:allObs} are taken from \citet{herczeg2014optical} and include one star which is now believed to be later than M3.

\begin{figure}
\centering
\includegraphics[width=\columnwidth]{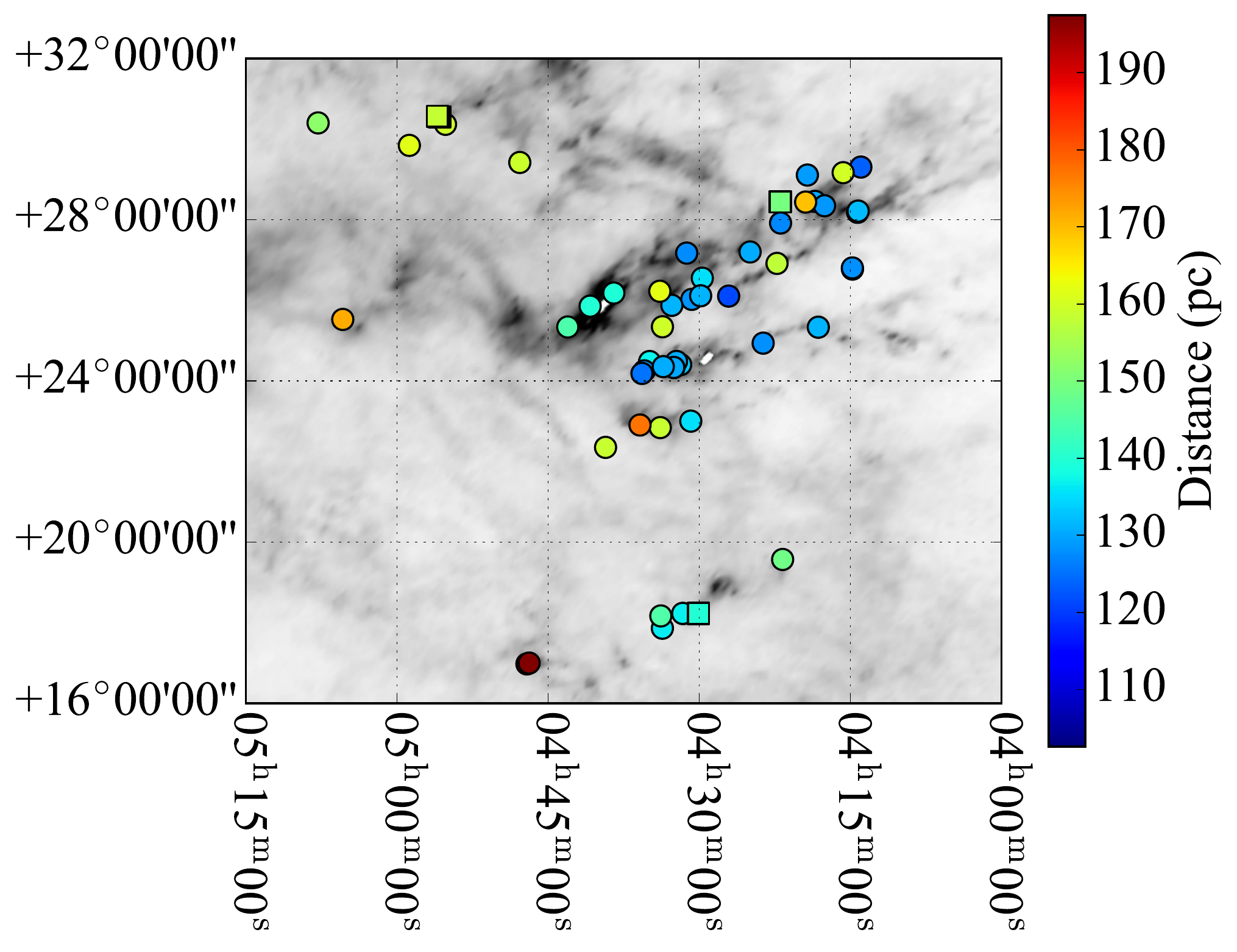}
\caption{Positions of our targets superimposed on the dust reddening map from \citet{schlafly2014map}. The squares represent the bright targets imaged with the PAH filter and circles are all other targets.}
\label{fig:taurusMap}
\end{figure}

\begin{figure}
\centering
\includegraphics[width=\columnwidth]{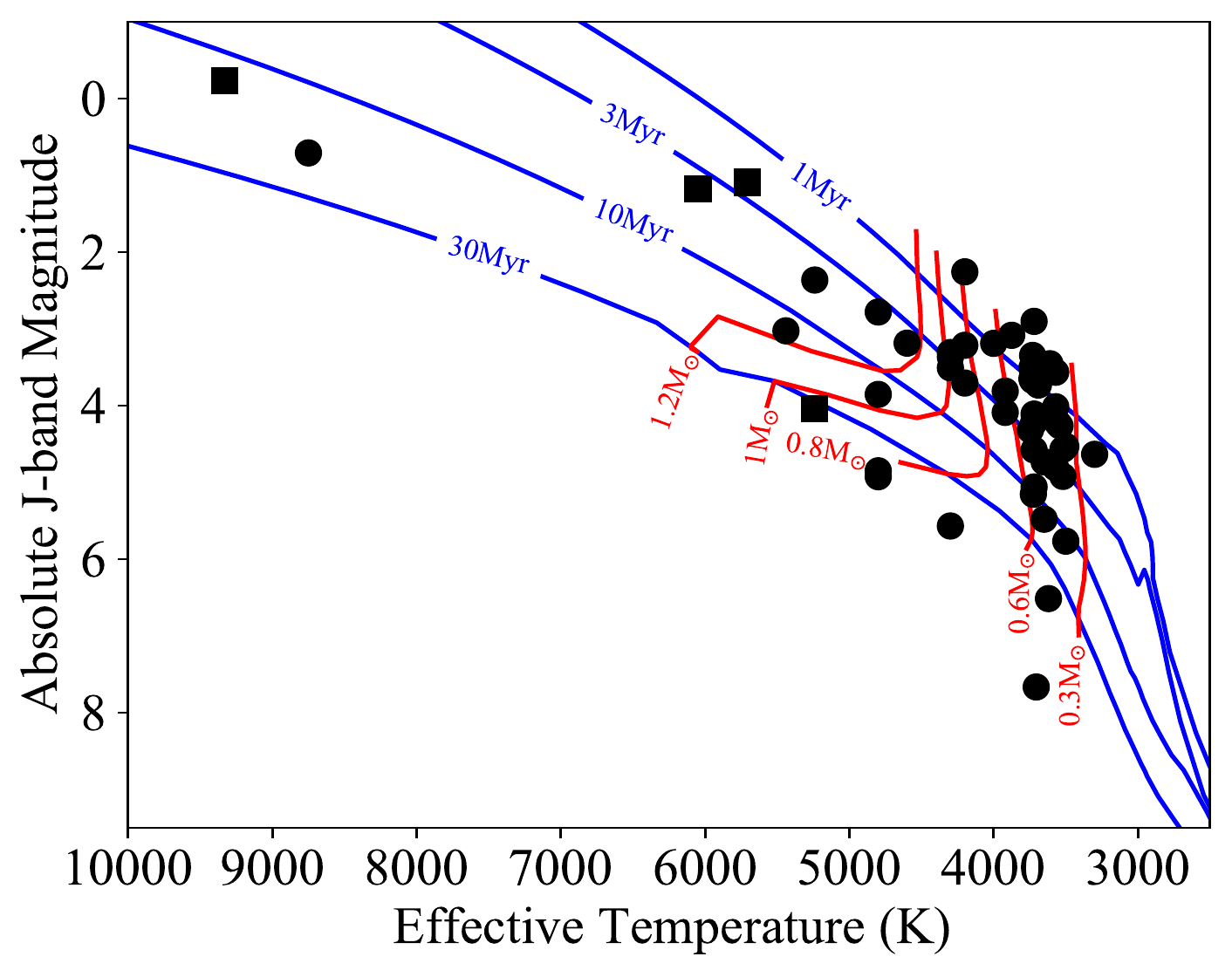}
\caption{H-R diagram of our targets plotting R-band magnitude against effective temperature.  As in Figure~\ref{fig:taurusMap}, The squares represent the bright targets imaged with the PAH filter and circles are all other targets.  MWC 480 and AB Aur are A-type stars and are the only stars in our sample hotter than 5000\,K, and hence they appear as outliers.}
\label{fig:HRdiagram}
\end{figure}

All of our targets were observed with an L' filter with the exception of RY Tau, AB Aur, UX Tau and SU Aur.  These stars were observed with a PAH filter as they are too bright for the method described in Section~\ref{sec:psf_subtraction} to work properly.  Our targets are mapped out in Figure~\ref{fig:taurusMap} and shown on an H-R diagram in Figure~\ref{fig:HRdiagram}.  The map in Figure~\ref{fig:taurusMap} also includes the distances taken from Gaia and a map of dust reddening from \citet{schlafly2014map}.  The H-R diagram in Figure~\ref{fig:HRdiagram} plots the absolute magnitude in the J-band (corrected for extinction using models from \citet{fitzpatrick2007analysis}) against effective temperature.  The effective temperature was calculated using spectral types from \citet{herczeg2014optical}.  Isochrones and isomass curves are shown using models from \citet{baraffe2015new}.  Note that several of our targets below 5000~K are under-luminous and appear older than 10\,Myr.  This is due to local reddening which is not taken into account and should not be regarded as the actual age of the star.

As shown in Figure~\ref{fig:taurusMap}, most of our targets are in the main region of the TMC at distances of 130--150\,pc.  There are some outliers, most notably DQ Tau and DR Tau at distances of $\sim$190\,pc, and separated from the main group.  The stellar properties for all of our targets are presented in Table~\ref{tab:allObs}.  The temperature was converted to mass using the evolutionary tracks from \citet{baraffe2015new}.
\begin{table*}
\caption{Properties of target stars in the TMC.  The spectral types come from \citet{herczeg2014optical}, the masses are calculated using evolutionary tracks from \citet{baraffe2015new} and the W$_{1}$ magnitudes are from the WISE catalog.  The distance comes from the Gaia DR2.}
\begin{tabular}{l|c|c|c|c|c|c|c|c}
Name & R.A. & Dec & Distance & Mass ($M_{\odot}$) & SpT & Rp (mag) & K (mag) & W$_{1}$ (mag)\\ \hline
IRAS 04108+2910 & 04 13 57.38 & +29 18 19.3 & 123.5$\pm$1.5 & 0.40 & M3.0 & 14.0 & 9.36 & 7.93\\
FM Tau & 04 14 13.58 & +28 12 49.2 & 131.9$\pm$0.8 & 0.24 & M4.5 & 12.5 & 8.76 & 8.00\\
CW Tau & 04 14 17.00 & +28 10 57.8 & 132.4$\pm$0.7 & 1.00 & K3.0 & 11.7 & 7.13 & 5.80\\
FP Tau & 04 14 47.31 & +26 46 26.4 & 128.5$\pm$0.9 & 0.36 & M2.6 & 11.6 & 8.87 & 8.40\\
CX Tau & 04 14 47.86 & +26 48 11.0 & 127.9$\pm$0.6 & 0.37 & M2.5 & 11.5 & 8.81 & 8.52\\
2MASS J04154278+2909597 & 04 15 42.79 & +29 09 59.8 & 160.0$\pm$1.7 & 0.51 & M0.6 & 12.8 & 9.38 & 9.04\\
CY Tau & 04 17 33.73 & +28 20 46.8 & 128.9$\pm$0.7 & 0.38 & M2.3 & 11.4 & 8.60 & 7.79\\
V409 Tau & 04 18 10.79 & +25 19 57.4 & 131.4$\pm$0.7 & 0.54 & M0.6 & 11.2 & 9.03 & 8.32\\
V410 Tau & 04 18 31.10 & +28 27 16.2 & 130.4$\pm$0.9 & 1.49 & K3.0 & 9.5 & 7.63 & 7.36\\
BP Tau & 04 19 15.83 & +29 06 26.9 & 129.1$\pm$1.0 & 0.45 & M0.5 & 10.6 & 7.74 & 7.11\\
V836 Tau & 04 19 26.27 & +28 26 14.3 & 169.6$\pm$1.2 & 0.45 & M0.8 & 11.6 & 8.60 & 8.19\\
IRAS 04187+1927 & 04 21 43.27 & +19 34 13.3 & 148.7$\pm$2.2 & 0.37 & M2.4 & 13.0 & 8.02 & 7.11\\
DE Tau & 04 21 55.63 & +27 55 06.2 & 127.4$\pm$1.1 & 0.37 & M2.3 & 10.8 & 7.80 & 7.08\\
RY Tau & 04 21 57.41 & +28 26 35.5 & 149.6$\pm$5.4 & 2.41 & G0.0 & 9.9 & 5.39 & 4.24\\
2MASS J04221675+2654570 & 04 22 16.76 & +26 54 57.1 & 157.6$\pm$3.4 & 0.52 & M1.5 & 14.8 & 9.01 & 7.73\\
FT Tau & 04 23 39.19 & +24 56 14.1 & 127.8$\pm$0.8 & 0.37 & M2.8 & 12.4 & 8.60 & 7.65\\
IP Tau & 04 24 57.08 & +27 11 56.5 & 130.6$\pm$0.7 & 0.47 & M0.6 & 11.4 & 8.35 & 7.71\\
DG Tau & 04 27 04.69 & +26 06 16.0 & 121.2$\pm$2.1 & 0.64 & K7.0 & 10.9 & 6.99 & 6.18\\
DH Tau & 04 29 41.56 & +26 32 58.3 & 135.4$\pm$1.3 & 0.37 & M2.3 & 11.4 & 8.18 & 7.40\\
IQ Tau & 04 29 51.56 & +26 06 44.9 & 131.3$\pm$1.1 & 0.43 & M1.1 & 12.0 & 7.78 & 7.27\\
UX Tau & 04 30 04.00 & +18 13 49.4 & 139.9$\pm$2.0 & 0.89 & K0.0 & 10.3 & 8.92 & 6.92\\
DK Tau & 04 30 44.25 & +26 01 24.5 & 128.5$\pm$1.0 & 0.52 & K8.5 & 11.3 & 7.10 & 6.12\\
IRAS 04278+2253 & 04 30 50.28 & +23 00 08.8 & 135.4$\pm$1.3 & 1.24 & G8.0 & 11.4 & 5.86 & 4.53\\
JH 56 & 04 31 14.44 & +27 10 17.9 & 127.5$\pm$0.7 & 0.64 & K8.0 & 11.0 & 8.79 & 8.74\\
LkHa 358 & 04 31 36.14 & +18 13 43.3 & 102.6$\pm$5.1 & 0.50 & M0.9 & 16.1 & 9.69 & 8.21\\
HL Tau & 04 31 38.44 & +18 13 57.7 & 136.7$\pm$2.2 & 0.79 & K3c & 15.7 & 7.41 & 5.24\\
HK Tau & 04 31 50.57 & +24 24 18.1 & 133.3$\pm$1.6 & 0.47 & M1.5 & 12.8 & 8.59 & 7.82\\
2MASS J04321540+2428597 & 04 32 15.41 & +24 28 59.8 & 130.5$\pm$3.2 & 0.67 & K5.5 & 13.6 & 8.10 & 6.62\\
FY Tau & 04 32 30.58 & +24 19 57.3 & 130.2$\pm$1.2 & 0.51 & M0.1 & 12.5 & 8.05 & 7.32\\
FZ Tau & 04 32 31.76 & +24 20 03.0 & 130.0$\pm$1.3 & 0.50 & M0.5 & 12.5 & 7.35 & 6.15\\
UZ Tau A & 04 32 42.88 & +25 52 31.9 & 131.2$\pm$1.6 & 0.39 & M1.9 & 11.2 & 7.35 & 6.25\\
GI Tau & 04 33 34.06 & +24 21 17.1 & 130.5$\pm$0.8 & 0.46 & M0.4 & 11.5 & 7.89 & 7.09\\
DL Tau & 04 33 39.08 & +25 20 38.1 & 159.3$\pm$1.2 & 0.99 & K5.5 & 11.1 & 7.96 & 6.94\\
HN Tau A & 04 33 39.35 & +17 51 52.4 & 136.6$\pm$2.9 & 0.79 & K3c & 12.5 & 8.38 & 7.23\\
DM Tau & 04 33 48.73 & +18 10 10.0 & 145.1$\pm$1.1 & 0.35 & M3.0 & 12.0 & 9.52 & 9.46\\
CI Tau & 04 33 52.01 & +22 50 30.1 & 158.7$\pm$1.2 & 0.97 & K5.5 & 11.1 & 7.79 & 6.76\\
IT Tau & 04 33 54.70 & +26 13 27.5 & 162.0$\pm$2.0 & 0.89 & K6.0 & 12.0 & 7.86 & 7.40\\
AA Tau & 04 34 55.42 & +24 28 53.2 & 137.2$\pm$2.4 & 0.45 & M0.6 & 13.1 & 8.05 & 7.45\\
DN Tau & 04 35 27.38 & +24 14 58.9 & 128.2$\pm$0.9 & 0.46 & M0.3 & 10.5 & 8.02 & 7.66\\
2MASS J04354093+2411087 & 04 35 40.94 & +24 11 08.8 & 125.2$\pm$2.3 & 0.58 & M0.5 & 14.2 & 8.41 & 7.35\\
HP Tau & 04 35 52.78 & +22 54 23.2 & 177.1$\pm$3.4 & 1.26 & K4.0 & 11.8 & 7.62 & 6.02\\
DO Tau & 04 38 28.58 & +26 10 49.4 & 139.4$\pm$1.0 & 0.46 & M0.3 & 11.3 & 7.30 & 6.34\\
LkCa 15 & 04 39 17.79 & +22 21 03.4 & 158.9$\pm$1.2 & 0.97 & K5.5 & 10.7 & 8.16 & 7.50\\
JH 223 & 04 40 49.51 & +25 51 19.2 & 139.9$\pm$1.1 & 0.38 & M2.8 & 12.9 & 9.49 & 8.94\\
GO Tau & 04 43 03.08 & +25 20 18.7 & 144.6$\pm$1.0 & 0.41 & M2.3 & 12.6 & 9.33 & 8.97\\
DQ Tau & 04 46 53.06 & +17 00 00.1 & 197.4$\pm$2.0 & 0.43 & M0.6 & 11.4 & 7.98 & 7.09\\
DR Tau & 04 47 06.21 & +16 58 42.8 & 195.7$\pm$2.5 & 0.78 & K6.0 & 10.7 & 6.87 & 5.83\\
DS Tau & 04 47 48.60 & +29 25 11.2 & 159.1$\pm$1.1 & 0.46 & M0.4 & 11.1 & 8.04 & 7.35\\
GM Aur & 04 55 10.98 & +30 21 59.5 & 159.6$\pm$2.1 & 0.84 & K6.0 & 10.8 & 8.28 & 8.30\\
AB Aur & 04 55 45.85 & +30 33 04.3 & 162.9$\pm$1.5 & 1.84 & A1.0 & 6.9 & 4.23 & 3.25\\
SU Aur & 04 55 59.39 & +30 34 01.5 & 158.4$\pm$1.5 & 2.65 & G4.0 & 8.8 & 5.99 & 5.07\\
MWC 480 & 04 58 46.26 & +29 50 37.0 & 161.8$\pm$2.0 & 2.03 & A3.0 & 7.5 & 5.53 & 4.87\\
2MASS J05052286+2531312 & 05 05 22.86 & +25 31 31.2 & 171.9$\pm$2.6 & 0.45 & M1.8 & 14.1 & 11.16 & 9.17\\
RW Aur A & 05 07 49.76 & +30 24 03.7 & 151.9$\pm$20.8 & 1.77 & K0c & 11.4 & 7.02 & 6.25\\
V819 Tau & 05 16 22.30 & +27 26 24.2 & 131.7$\pm$1.1 & 0.61 & K8.0 & 11.1 & 8.42 & 8.27\\
\end{tabular}
\label{tab:allObs}
\end{table*}

\section{Observations and Image Analysis}
\label{sec:observations_analysis}
\subsection{Observations}
\label{sec:observations}
Our observations were made using the NIRC2 camera of the Keck II telescope on 27, 28 November, 5 December 2015 and 7,8,9 November 2016. As the focus of these observations was to search for close companions, we used the 512x512 sub-array mode in order to minimise overheads - noting that the readout time would have often decreased our duty cycle by a factor of $\sim 2$ had we used the full array \footnote{https://www2.keck.hawaii.edu/inst/nirc2/ObserversManual.html}.\\

In order to account for irregularities in the telescope PSF, at least 2~position angles were required for each object. Weather permitting, every object was observed in 4~observing blocks: 2~in the first half of the night and~2~in the second half. Where possible we avoided the highest elevations where azimuth slew rates are high and telescope vibrations can affect observations. Based on past experience with Keck, our objects were divided into groups of~4~and observed in the following sequence: A,B,C,D,~A,B,C,D which gave us 2~observations of 4~objects.  The members of the group are determined by their proximity to each other. This sequence is then repeated in the second half of the night. Each observation consisted of a number of frames (usually~6) with average exposure times of 30\,s, which is composed of a small integration time multiplied by an appropriate number of co-adds (snapshots that make up the final image), which also varies depending on the brightness of the target. A summary of our observations is shown in Table~\ref{tab:obs}.
\begin{table*}
\caption{Details of observations.  T$_\mathrm{int}$ refers to the integration time for each co-add. This is multiplied by the number of co-adds to get the exposure time T$_\mathrm{exp}$ for each frame. The \# of visits column gives the number of observing blocks taken for that object each night. The number of values in this column is the number of observing nights. The \# of frames column shows the number of frames for each observing block in the order they were taken. For example, IRAS~04108+2910 was observed on only one night and visited 4~times with 6~frames in each block.  FM~Tau was observed on 2~nights with 3~visits on the first night, and~1~on the second. The blocks taken on the first night had 12, 12 and 6~frames, while the block taken on the second night had~6.}
\begin{tabular}{l|l|l|l|l|l|l}
Name & Obs. Date & T$_\mathrm{int}$(s) & Coadds & T$_\mathrm{exp}$ (s) & \# Visits & \# Frames\\\hline
IRAS 04108+2910 & 2016-11-09 & 0.4 & 80 & 32 & 4 & 6,6,6,6\\
FM Tau & 2015-11-27,2015-11-28 & 0.3 & 100 & 30 & 3,1 & 12,12,6,6\\
CW Tau & 2016-11-07 & 0.2 & 160 & 32 & 4 & 6,7,6,6\\
FP Tau & 2016-11-07,2016-11-09 & 0.1 & 320 & 32 & 4,1 & 6,6,6,6,6\\
CX Tau & 2016-11-07,2016-11-09 & 0.4 & 80 & 32 & 5,1 & 6,6,6,6,6,6\\
2MASS J04154278+2909597 & 2016-11-09 & 0.4 & 80 & 32 & 4 & 7,6,6,6\\
CY Tau & 2015-11-27,2015-11-28 & 0.3 & 100 & 30 & 3,2 & 12,12,6,6,6\\
V409 Tau & 2016-11-09 & 0.4 & 80 & 32 & 4 & 7,6,6,6\\
V410 Tau & 2015-11-28,2016-11-07 & 0.2 & 160 & 32 & 1,1 & 16,6\\
BP Tau & 2015-11-27,2015-11-28 & 0.15 & 200 & 30 & 3,1 & 12,12,6,5\\
V836 Tau & 2015-11-27,2015-11-28 & 0.3 & 100 & 30 & 4,1 & 6,6,6,6,6\\
IRAS 04187+1927 & 2016-11-08 & 0.4 & 80 & 32 & 4 & 6,10,8,6\\
DE Tau & 2015-11-27,2015-11-28,2015-12-05 & 0.3 & 100 & 30 & 2,2,2 & 6,8,6,6,8,6\\
RY Tau & 2015-11-27,2015-11-28 & 1.0 & 30 & 30 & 1,4 & 12,7,6,6,6\\
2MASS J04221675+2654570 & 2016-11-08 & 0.4 & 80 & 32 & 4 & 9,6,6,6\\
FT Tau & 2016-11-09 & 0.4 & 80 & 32 & 4 & 6,6,6,6\\
IP Tau & 2015-11-27,2015-11-28,2015-12-05 & 0.3 & 100 & 30 & 3,1,1 & 12,12,6,10,6\\
DG Tau & 2015-11-27,2015-11-28,2015-12-05 & 0.15 & 200 & 30 & 2,1,3 & 6,6,6,15,4,6\\
DH Tau & 2016-11-08 & 0.4 & 80 & 32 & 4 & 6,11,6,6\\
IQ Tau & 2015-11-27,2015-11-28 & 0.15 & 200 & 30 & 3,1 & 12,12,6,6\\
UX Tau & 2015-11-28 & 1.0 & 30 & 30 & 5 & 10,6,6,6,6\\
DK Tau & 2015-11-27,2015-11-28,2015-12-05,2016-11-08 & 0.15 & 200 & 30 & 2,2,2,4 & 6,8,6,6,6,8,6,6,6,6\\
IRAS 04278+2253 & 2016-11-08 & 0.053 & 600 & 31 & 1 & 7\\
JH 56 & 2015-11-27,2015-11-28 & 0.3 & 100 & 30 & 3,1 & 12,12,6,6\\
LkHa 358 & 2016-11-08,2016-11-09 & 0.4 & 80 & 32 & 4,1 & 6,4,6,6,6\\
HL Tau & 2016-11-07,2016-11-09 & 0.1 & 320 & 32 & 5,1 & 6,6,6,6,14,6\\
HK Tau & 2015-11-27,2016-11-09 & 0.3 & 100 & 30 & 1,3 & 9,8,6,6\\
2MASS J04321540+2428597 & 2016-11-07,2016-11-09 & 0.2 & 160 & 32 & 2,2 & 6,10,6,6\\
FY Tau & 2016-11-08,2016-11-09 & 0.4 & 80 & 32 & 3,1 & 6,6,6,6\\
FZ Tau & 2016-11-07 & 0.1 & 320 & 32 & 4 & 5,6,4,6\\
UZ Tau A & 2016-11-09 & 0.1 & 320 & 32 & 2 & 6,6\\
GI Tau & 2015-11-27,2015-11-28,2015-12-05 & 0.15 & 200 & 30 & 3,1,2 & 12,12,6,6,6,4\\
DL Tau & 2015-11-27,2015-11-28,2015-12-05 & 0.15 & 200 & 30 & 2,2,1 & 6,6,6,6,6\\
HN Tau A & 2016-11-08,2016-11-09 & 0.4 & 80 & 32 & 3,1 & 6,6,6,6\\
DM Tau & 2015-11-27,2015-11-28 & 0.3 & 100 & 30 & 2,3 & 6,6,12,12,12\\
CI Tau & 2015-11-27,2015-11-28 & 0.15 & 200 & 30 & 3,1 & 12,12,6,6\\
IT Tau & 2015-11-28 & 0.2 & 160 & 32 & 4 & 6,6,6,6\\
AA Tau & 2015-11-27,2016-11-09 & 0.4 & 80 & 32 & 1,4 & 12,9,6,5,4\\
DN Tau & 2015-11-28,2016-11-09 & 0.2 & 160 & 32 & 3,1 & 6,6,6,6\\
2MASS J04354093+2411087 & 2016-11-08,2016-11-09 & 0.4 & 80 & 32 & 3,1 & 6,6,6,6\\
HP Tau & 2015-11-27,2015-11-28,2015-12-05 & 0.15 & 200 & 30 & 3,1,2 & 12,12,6,6,6,10\\
DO Tau & 2016-11-07,2016-11-08 & 0.2 & 160 & 32 & 4,1 & 6,6,10,6,6\\
LkCa 15 & 2015-11-27,2015-11-28,2016-11-08 & 0.3 & 100 & 30 & 3,2,4 & 12,12,6,6,6,6,6,6,6\\
JH 223 & 2016-11-07,2016-11-08,2016-11-09 & 0.4 & 80 & 32 & 2,2,2 & 6,6,9,6,6,6\\
GO Tau & 2016-11-07,2016-11-08,2016-11-09 & 0.4 & 80 & 32 & 2,1,1 & 6,6,6,6\\
DQ Tau & 2015-11-27,2015-11-28 & 0.15 & 200 & 30 & 3,2 & 12,8,6,10,8\\
DR Tau & 2015-11-27,2016-11-09 & 0.15 & 200 & 30 & 2,3 & 12,6,6,6,6\\
DS Tau & 2015-11-27,2015-11-28 & 0.3 & 100 & 30 & 3,1 & 6,6,6,6\\
GM Aur & 2015-11-27,2015-11-28,2016-11-08 & 0.3 & 100 & 30 & 4,2,1 & 6,6,6,6,12,12,6\\
AB Aur & 2015-11-28 & 0.2 & 160 & 32 & 4 & 6,6,6,6\\
SU Aur & 2015-11-28 & 1.0 & 30 & 30 & 5 & 6,6,6,6,6\\
MWC 480 & 2016-11-07,2016-11-09 & 0.053 & 600 & 31 & 4,3 & 6,6,6,6,6,6,6\\
2MASS J05052286+2531312 & 2016-11-08,2016-11-09 & 0.4 & 80 & 32 & 2,2 & 8,6,6,6\\
RW Aur A & 2016-11-07,2016-11-09 & 1.0 & 30 & 30 & 3,4 & 8,6,5,6,6,6,6\\
V819 Tau & 2015-11-27,2015-11-28 & 0.3 & 100 & 30 & 3,1 & 12,12,6,6\\
\end{tabular}
\label{tab:obs}
\end{table*}

\subsection{Data Reduction}
Starting with our raw 512$\times$512 pixel images, we first subtracted the master dark frame for the night and divided by the flat frame.  For each observing block, the target was observed in two different dither positions.  Half the images had the target in the top left quarter and the other half had it in the bottom right corner.  This allowed us to calculate an approximate sky background for each image.  First, each image was cropped to a size of 192$\times$192 pixels (1.92''$\times$1.92'') which was centred on the star by calculating the peak of the image after applying a median filter and performing a simple pixel roll.  The corresponding area from the other dither position served as the sky background which was then subtracted from the cropped image.\\

Any ``bad'' pixels were fixed using the algorithm from \citet{ireland2013phase}.  Once identified, these pixels were set to the corresponding value in the median filtered image.  In addition to bad pixels identified in the dark and flat field imates, we also corrected with the same algorithm pixels near saturation  which were defined as any with counts greater than 17500\,$\times$ the number of coadds for that image.  The threshold of 17500 was chosen empirically to produce final PSF-subtracted images with the lowest residuals.  Pixels above this threshold were treated as bad pixels.  Once all images had been ``cleaned'' in this way, they were stored in a data cube containing all images for a particular observing block.  The analysis was then performed on these cleaned images.
\subsection{Image Analysis using PSF subtraction} \label{sec:psf_subtraction}

Our first method of image analysis is a form of reference star differential imaging (RDI) which focuses on simply removing the effect of the central star in order to look for planets.  To achieve this, we first went back to the basics of how an image is created.  When the telescope's optical system is applied, we assumed the signal from a planet will look the same as a star but reduced by a contrast ratio.  In other words, the star was represented by a PSF given by the properties of the optical system and the planet was represented by the same PSF but scaled by a contrast ratio and shifted by the planet's relative position.  In 1-D, this image function is given by
\begin{equation}
    i(x) = p(x)+cp(x-x_{0}),
    \label{eq:image}
\end{equation}
where $p$ is the PSF representing a single star, $c$ is the contrast ratio between a planet and the star and $x$ is a spatial variable with the star at $x=0$ and the planet at $x=x_{0}$.
 The first step in our analysis is the subtraction of the PSF.

For our PSF, we simply used the (cleaned) image of another star, which was taken at a similar time to our target.  Another possible approach would be principal component analysis (PCA) in which the PSF is taken from a linear combination of stellar images. The number of components in the analysis is optimised, which has shown promising results in reducing background noise and finding planets \citep{meshkat2013optimized,hunziker2018pca}. An extreme approach would be to create a linear combination using all our images. We have tried this approach as well as an optimisation and found that there was no significant improvement in our signal-to-noise ratio, so we do not report on this here. Instead, we have opted for the opposite extreme, in which we only use 1~image that is selected by optimisation.  Due to fewer degrees of freedom, this approach also subtracts a smaller fraction of the flux of a real companion than PCA.

\begin{figure*}
\subfigure[Cleaned Image of AA Tau]{\label{fig:AATauIm}\includegraphics[width=5.9cm]{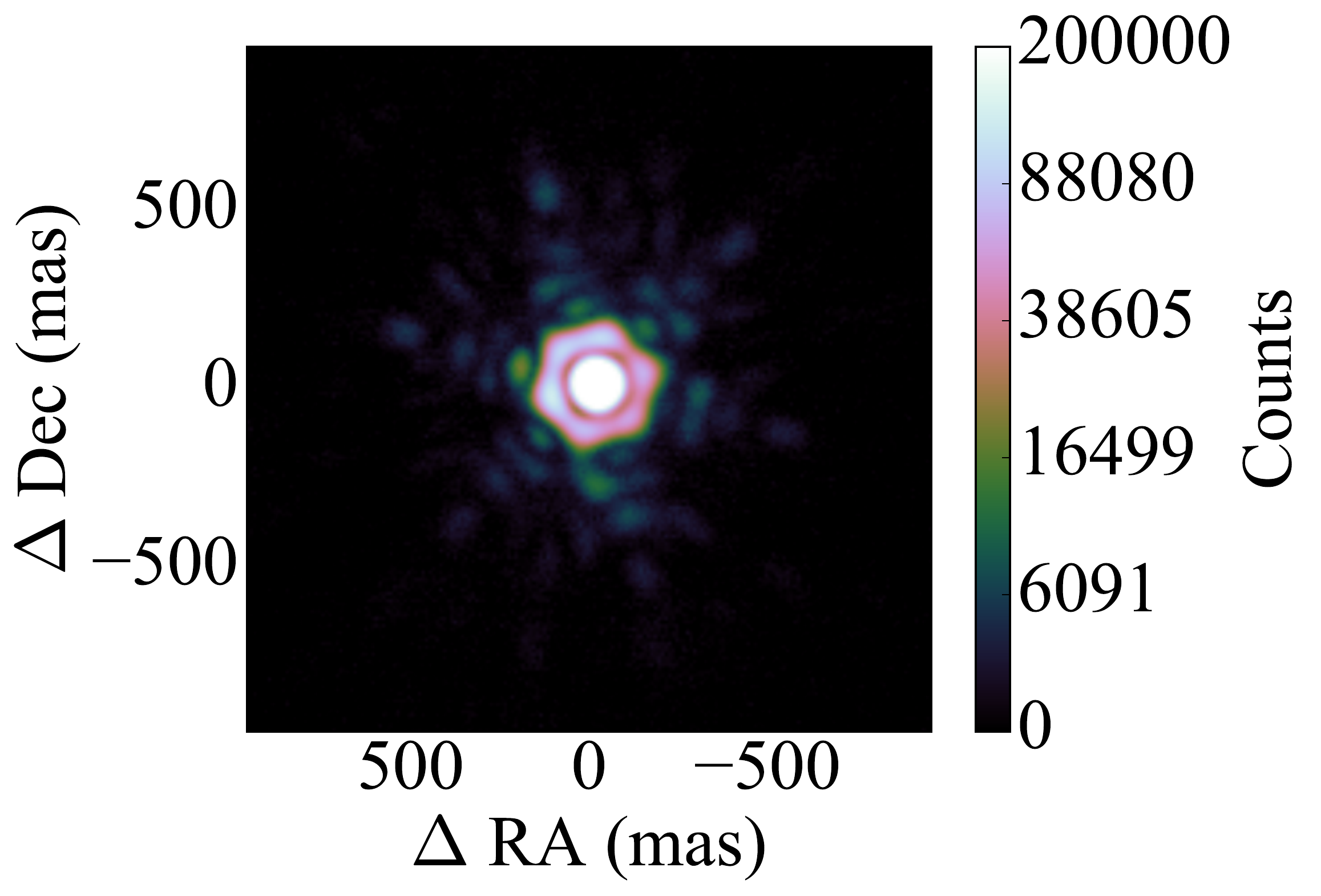}}
\subfigure[Difference Image]{\label{fig:AATauDiff}\includegraphics[width=5.7cm]{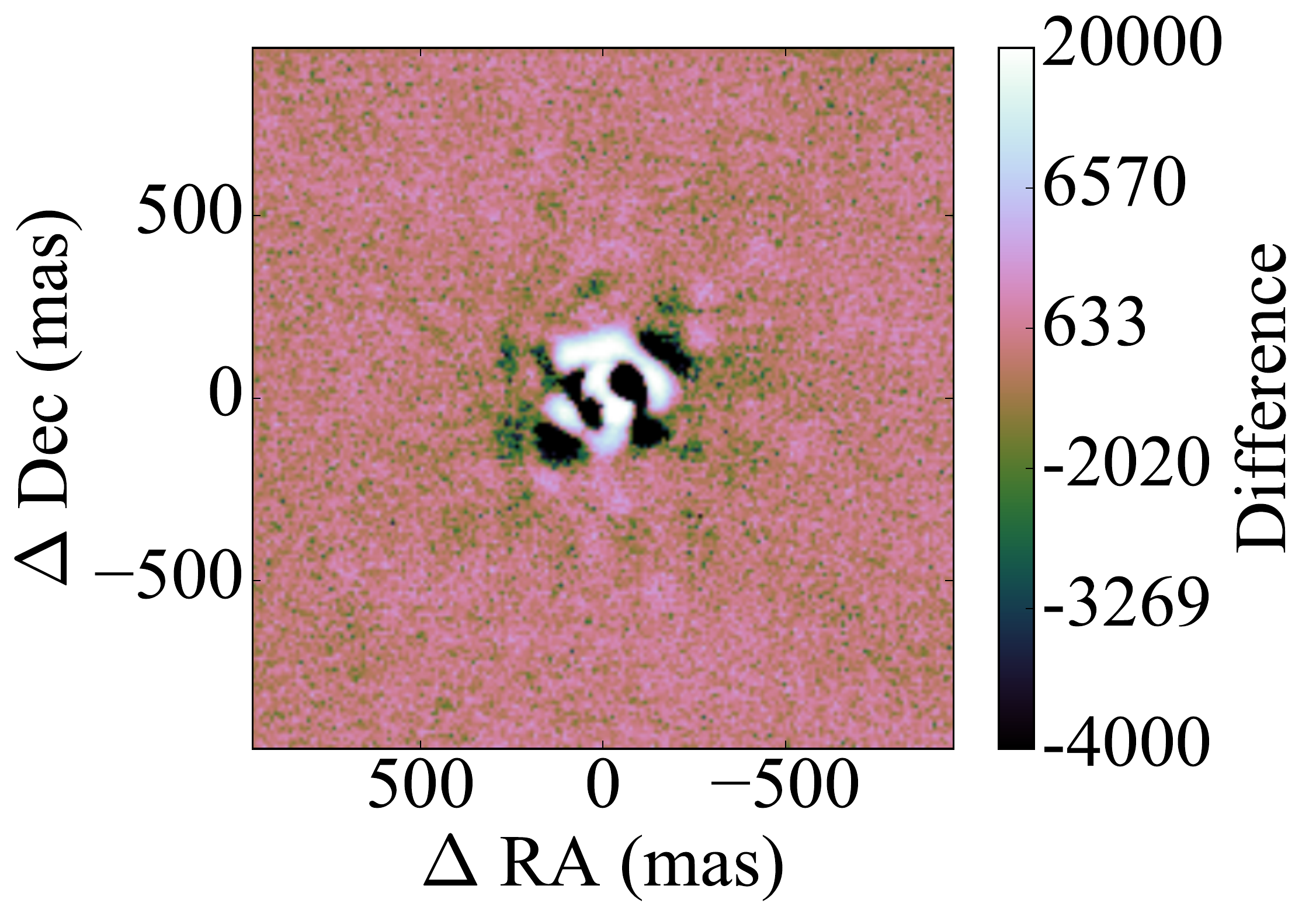}}
\subfigure[Contrast Map]{\label{fig:AATauCont}\includegraphics[width=5.7cm]{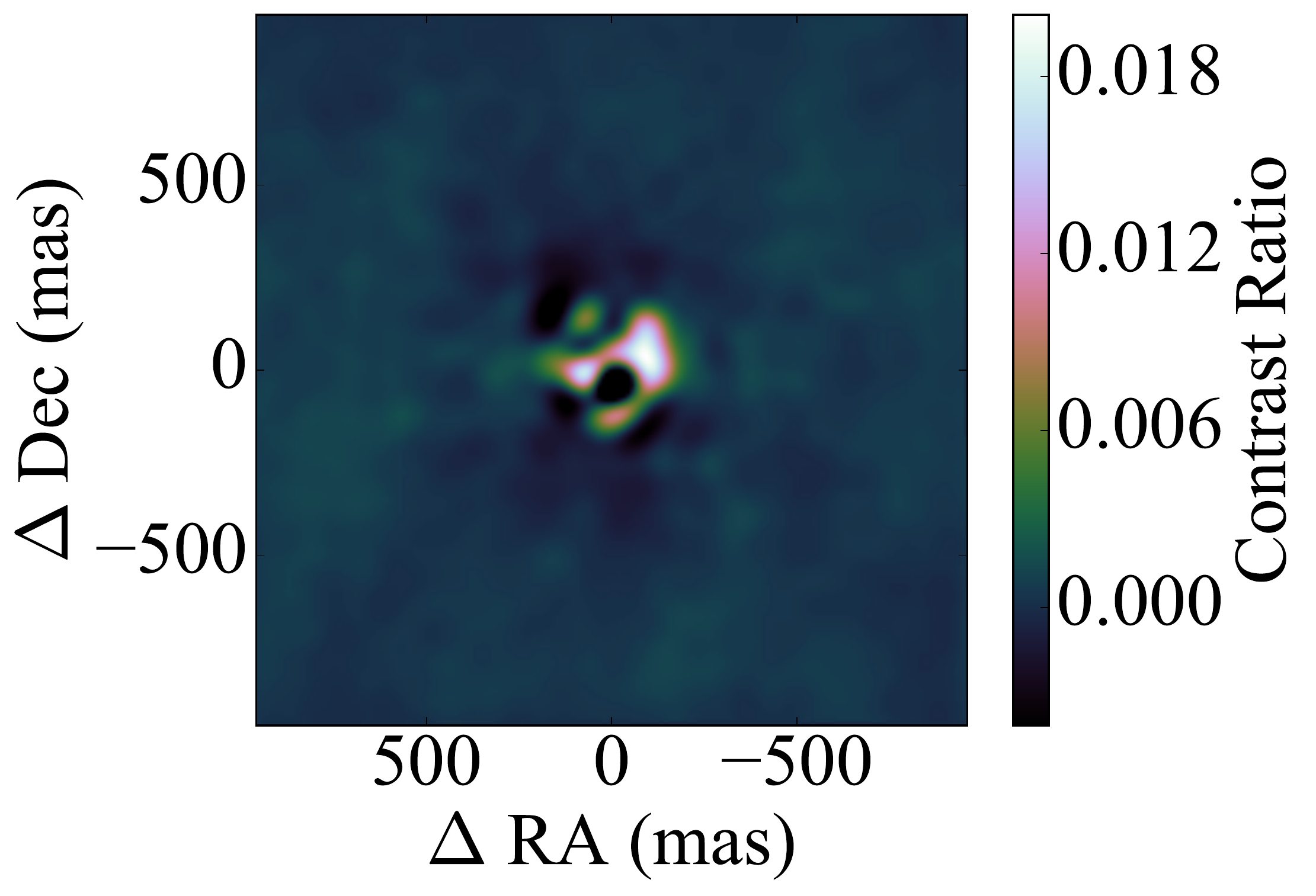}}
\caption{Example showing a reduced image of AA Tau \textbf{(left)}, the difference after an image of HK Tau is subtracted \textbf{(centre)} and the contrast ratio map of AA Tau \textbf{(right)}.  When the contrast ratio is negative, this is due to positive features in the subtracted image.  The bright feature in the middle of the contrast map is due to offsets in the position of the target and PSF central star.  At wider separations, we can use this method to search for companions.  While there are no obvious features in this example, the contrast values show that we should be able to detect companions more than 0.002 the brightness of the star.}
\label{fig:analysisEx}
\end{figure*}

Our targets are observed in blocks, typically consisting of 6 images each.  Every target image is matched with the image of another star, which plays the role of our PSF.  The PSF image for each target image was chosen from a selection of nearby observing blocks.  For a given target image, we have a set of potential PSFs $p_{n}$.  For each of these, we calculate the sum of the square of the differences given by
\begin{equation}
    \Sigma_{n} = \sum\limits_{ij}\left(t_{ij}-f_{n}p_{n,ij}\right)^{2},
\end{equation}
where $t$ is the target image, $p_{n}$ is the image of another star, which we use as the PSF, and $i$ and $j$ are pixel indices.  The scaling factor $f_{n}$ was chosen such that the target and PSF had the same maximum value so when they are subtracted, the central star cancels out.  For a given PSF this is given by
\begin{equation}
    f_{n} = \frac{\max(t)}{\max(p_{n})}.
\end{equation}
This is calculated for all possible PSF images $p_{n}$ in our sample, simply based on photon count.  Whichever produces the smallest value of $\Sigma_{n}$ is chosen as our PSF.  When this PSF is chosen, we then calculate the difference between the target and the PSF that has been multiplied by the scaling factor $f_{n}$.  Following on from Equation~\ref{eq:image}, the difference is represented in 1-D by
\begin{equation}
    d(x) = i(x) - p(x) = cp(x-x_{0}).
\end{equation}
We can then calculate a smooth contrast ratio as a function of position by cross-correlating the difference function with the PSF.  This is then divided by the PSF cross-correlated with itself in order to normalise the contrast ratio.  The contrast ratio as a function of position is given by
\begin{equation}
    c(x_{0}) = \frac{(d\star p)(x_{0})}{(p\star p)(x_{0})},
\end{equation}
where $\star$ denotes the cross-correlation operator given by
\begin{equation}
    (d\star p)(x_{0}) = \int\limits^{\infty}_{-\infty}d(x)\,p(x+x_{0})\, dx.
\end{equation}
When we apply this method to the image of one of our targets, this produces a map of the contrast ratio between any features and the central star.  An example is shown in Figure~\ref{fig:analysisEx} with an image of AA Tau.  This star was chosen simply because its properties are close to the average of our sample.  For the PSF, we used an image of HK Tau which, as shown in Table~\ref{tab:obs}, was taken on the same night.

This process is repeated for all images of the target, and the contrast maps are averaged.  By taking the average contrast about an annulus of fixed radius, we then produce a 1-D plot of the contrast limit against separation.

\subsection{Kernel phase data reduction}
\label{sec:kernel_phase_data_reduction}
Complementary to the PSF subtraction (cf., Section~\ref{sec:psf_subtraction}) we use the kernel phase technique in order to search for companions close to the host star, inside of $500~\text{mas}$. This analysis begins with the same 192 x 192 pixel cleaned data cubes described in Section~\ref{sec:psf_subtraction}.

The kernel phase technique finds a special linear combination of the Fourier phase $\phi$ (i.e., the phase of the Fourier transform of the image) which is independent of pupil plane phase $\varphi$ (phase aberrations in the telescope pupil which cause quasi-static speckles) to second order, similar to closure phase in non-redundant masking, but for full pupil images (i.e., highly redundant apertures). Let $\bm{A}$ be the baseline-mapping matrix introduced by \citet{martinache2010}, which maps the sub-apertures in the pupil plane (cf., left panel of Figure~\ref{fig:pupil_model}) to their corresponding Fourier plane baselines (cf., right panel of Figure~\ref{fig:pupil_model}), then the Fourier phase $\phi$ observed through the telescope is
\begin{equation}
    \phi = \bm{R}^{-1}\cdot\bm{A}\cdot\varphi+\phi_\text{obj}+\mathcal{O}(\varphi^3),
    \label{eq:phase}
\end{equation}
where $\bm{R}$ encodes the redundancy of the Fourier plane baselines and $\phi_\text{obj}$ is the phase intrinsic to the observed astronomical object (which is the quantity that we would like to measure). This problem is significantly simplified by multiplying Equation~\ref{eq:phase} with the kernel $\bm{K}$ of $\bm{R}^{-1}\cdot\bm{A}$, i.e.,
\begin{equation}
    \theta = \bm{K}\cdot\phi = \underbrace{\bm{K}\cdot\bm{R}^{-1}\cdot\bm{A}}_{=0}\cdot\varphi+\bm{K}\cdot\phi_\text{obj}+\mathcal{O}(\varphi^3),
\end{equation}
so that the kernel phase observed through the telescope $\theta$ is directly equal to the kernel phase intrinsic to the observed object $\theta_\text{obj} = \bm{K}\cdot\phi_\text{obj}$ (except for higher order noise terms).

The kernel phase technique was first used by \citet{martinache2010} who demonstrated the detection of a 10:1 companion at $0.5~\lambda/D$ in HST/NICMOS data, clearly showing the improved speckle calibration capabilities with respect to image plane data reduction techniques. More recently, \citet{pope2016} applied kernel phase to ground-based observations of $\alpha$ Oph with the 5.1 m Hale Telescope and showed that it outperforms PSF fitting and bispectral analysis under appropriate conditions (i.e., high Strehl). \citet{kammerer2019kernel} further developed the technique including a principal component calibration based on Karhunen-Lo\`eve decomposition \citep{soummer2012} for the subtraction of the residual kernel phase signal measured on calibrator stars and detected eight (candidate) low-mass stellar companions (five of which were previously unknown) in an archival VLT/NACO data set, one of which is separated by only $0.8~\lambda/D$.

Here, we use the same kernel phase data reduction pipeline as \citet{kammerer2019kernel}, with slight modifications and improvements explained below.

\subsubsection{Kernel phase extraction}
\label{sec:kernel_phase_extraction}

\begin{figure}
\centering
\includegraphics[width=\columnwidth]{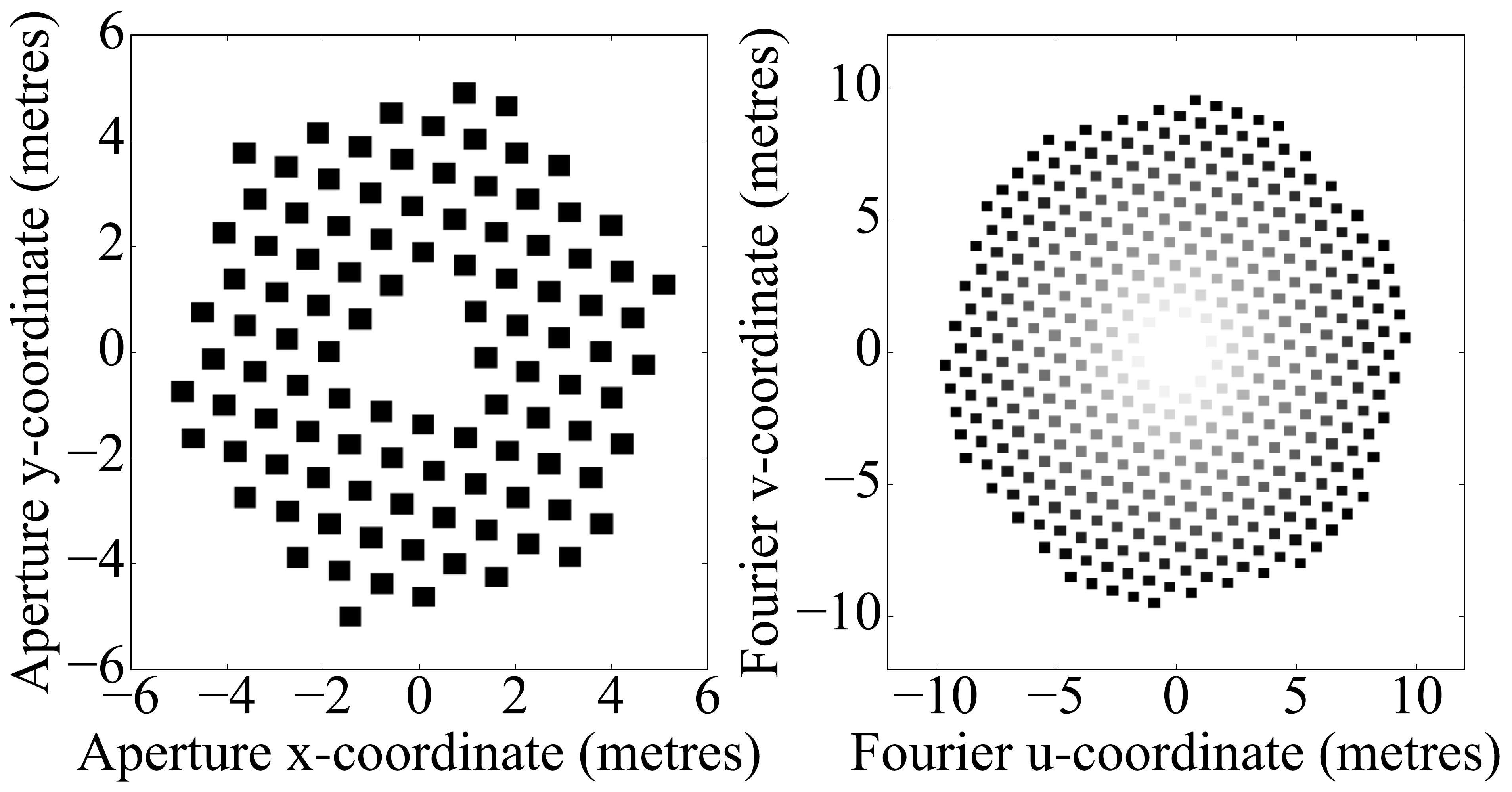}
\caption{Keck pupil model consisting of 105 individual sub-apertures (left panel) and its Fourier plane coverage with 205 distinct baselines (right panel) resulting in 100 individual kernel phases. The shading in the right panel shows the redundancy (multiplicity) of the baselines with dark representing low redundancy and bright representing high redundancy. Note that the right panel is the auto-correlation of the left panel \citep[cf., Section~2.1 of][]{martinache2010}.}
\label{fig:pupil_model}
\end{figure}

For extracting the kernel phase from the images we use the Python library XARA\footnote{\url{https://github.com/fmartinache/xara}}. XARA windows the cleaned images with a super-Gaussian mask, applies a linear discrete Fourier transform to them and performs a sub-pixel re-centring directly in the complex visibility space afterwards. Then, the Fourier phase $\phi$ of the images is extracted and multiplied by the kernel $\bm{K}$ of the transfer matrix $\bm{R}^{-1}\cdot\bm{A}$ of our Keck pupil model (Figure~\ref{fig:pupil_model}) yielding the kernel phase $\theta$ of the images \citep[cf., Section~2.1 of][]{martinache2010}.

For the super-Gaussian mask we use a radius of 50~pixels (i.e., $500~\text{mas}$) or a FWHM of 100~pixels. Our Keck pupil model consists of three individual sub-apertures per hexagonal Keck primary mirror segment in order to be sensitive to the tip-tilt orientation of each segment. Those sub-apertures that are behind the central obscuration from the secondary mirror are simply ignored. The sub-apertures are distributed uniformly in the plane of the primary mirror with a spacing of $b_\text{min} = 0.9~\text{m}$ resulting in a field of view of $\lambda/b_\text{min} \approx 865~\text{mas}$ and a maximum baseline of $b_\text{max} = 9.5~\text{m}$, yielding a resolution of $\sim\lambda/(2b_\text{max}) \approx 40~\text{mas}$. However, since the super-Gaussian mask has a radius of only $500~\text{mas}$, we restrict our search for companions (with the kernel phase technique) to angular separations of $40~\text{mas} \leq \rho \leq 500~\text{mas}$.

\subsubsection{Kernel phase frame selection}
\label{sec:kernel_phase_frame_selection}

Before we feed the kernel phase extracted from the images into our calibration and model fitting pipeline (cf., Section~\ref{sec:kernel_phase_calibration} and~\ref{sec:kernel_phase_model_fitting}), we perform a frame selection based on the sum of the squared kernel phase of each image, i.e.,
\begin{equation}
    \text{SOSK} = \sum_i|\theta_i^2|.
\end{equation}
From each night, we only keep the 50\% best images in the set of potential calibrators and the 75\% best images in the set of potential targets, where best means smallest $\text{SOSK}$. This is motivated by the fact that a point-symmetric source (e.g., a single star) has zero Fourier phase $\phi$ and therefore zero kernel phase $\theta$. Hence, a single star with a faint companion should still have a small kernel phase signal and images with a high kernel phase signal can usually be attributed to bad seeing conditions where the kernel phase technique is not valid (due to too much higher-order phase noise). Note that an unknown companion around one of our calibrators would have a small impact only, since we are averaging over a large number of calibrators and do not de-rotate them before we subtract them from the science target, so that the averaging is destructive in case of pupil-stabilised observations.

\subsubsection{Kernel phase calibration}
\label{sec:kernel_phase_calibration}

Similar to observations with an interferometer, we have to calibrate the kernel phase of our targets by subtracting the kernel phase of calibrators. This is done using the Karhunen-Lo\`eve projection described in Section~2.3 of \citet{kammerer2019kernel}. We perform the Karhunen-Lo\`eve calibration separately for each night since we found this to yield a smaller reduced $\chi^2$ than calibrating data from multiple nights together. The reason for this is likely that the quasi-static phase aberrations (for which we try to compensate with our calibration) are only stable over timescales of minutes to hours.

From the 75\% best images of each night, we select one object as a target and all images of different objects from the 50\% best images of the same night as calibrators.  Then, we subtract the first four Karhunen-Lo\`eve components from the kernel phase of the target $\theta$ and its uncertainties $\bm{\Sigma}_\theta$, i.e.,
\begin{align}
    \theta' &= \bm{P}'\cdot\theta, \\
    \bm{\Sigma}'_\theta &= \bm{P}'\cdot\bm{\Sigma}_\theta\cdot\bm{P}'^T,
\end{align}
where $\bm{\Sigma}_\theta$ and $\bm{P}'$ are obtained as described in Sections~2.2.3 and~2.3 of \citet{kammerer2019kernel}.

\subsubsection{Kernel phase model fitting}
\label{sec:kernel_phase_model_fitting}

After calibrating the kernel phase, we fit the binary model
\begin{equation}
    \theta'_\text{bin} = \bm{P}'\cdot\bm{K}\cdot\mathrm{arg}\left(1+c\exp\left(-2\pi i\left(\frac{\Delta_\text{RA}u}{\lambda}+\frac{\Delta_\text{DEC}v}{\lambda}\right)\right)\right),
\end{equation}
where $0 \leq c \leq 1$ is the companion contrast, $\Delta_\text{RA}$ and $\Delta_\text{DEC}$ are the on-sky separation of the companion, $u$ and $v$ are the Fourier coordinates of the pupil model and $\lambda = 3.776$\,$\mu$m is the observing wavelength, using a grid search and a least-squares routine as described in Section~2.4 of \citet{kammerer2019kernel} to it. We fit to all images of the same target simultaneously, also when a target was observed during multiple nights. Using the uncertainties $\bm{\Sigma}'_\theta$ derived from the photon noise of the images this yields a RA-DEC map of best-fit companion contrasts $\bm{c}_\text{fit}$ and their uncertainties $\bm{\sigma}_{c_\text{fit}}$ whose ratio is the photon noise-based signal-to-noise ratio $\text{SNR}_\text{ph}$. The grid position with the smallest reduced $\chi^2$ (obtained from a least-squares routine) is our best-fit companion.

\subsubsection{Empirical kernel phase detection limits}
\label{sec:empirical_kernel_phase_detection_limit}

If the uncertainties $\bm{\Sigma}_\theta$ derived from the photon noise would describe the underlying errors correctly (i.e., if all other errors would be negligible) we could simply classify those best-fit companions whose $\text{SNR}_\text{ph} > 5$ as significant detections. However, although readout noise and dark current are negligible for our dataset, there is a lot of higher-order phase noise, which leads to a high $\text{SNR}_\text{ph}$ and false detections for all of our targets (cf., column ``$\text{SNR}_\text{ph}$'' of Table~\ref{tab:kernel_phase_analysis}). Note that the kernel phase is independent of pupil plane phase noise only to second order and higher-order phase noise might be introduced by atmospheric turbulence or imperfect telescope optics.

Hence, an empirical method is necessary to derive robust detection limits. We classify the 1/3 of the targets with the highest $\text{SNR}_\text{scaled}$ as candidate detections and the rest of the targets as calibrators (cf., columns ``Can?'' and ``Cal?'' of Table~\ref{tab:kernel_phase_analysis}). Here, $\text{SNR}_\text{scaled}$ is the photon noise-based SNR scaled by the K-band magnitude of the object, i.e.,
\begin{equation}
    \text{SNR}_\text{scaled} = \text{SNR}_\text{ph}\sqrt{\frac{1}{10^{-(K-K_\text{med})/2.5}}},
\end{equation}
where $K_\text{med}$ is the median K-band magnitude of our targets. This scaling is motivated by the fact that the brighter objects have higher photon noise-based SNRs due to smaller uncertainties, but similar quasi-static errors. Then, we repeat the Karhunen-Lo\`eve calibration (only allowing images of objects in the list of calibrators to be selected as calibrators) and the model fitting. Afterwards, we compute an empirical detection limit $\bm{\sigma}_\text{emp}$ and an empirical signal-to-noise ratio,
\begin{equation}
    \text{SNR}_\text{emp} = \frac{\bm{c}_\text{fit}}{\bm{\sigma}_\text{emp}}
\end{equation}
for each of the candidate detections as described in Section~2.4.3 of \citet{kammerer2019kernel} and classify a candidate detection as significant, if $\text{SNR}_\text{emp} > 5$.  Note that this empirical detection limit is based on azimuthally averaging the contrast maps $\bm{c}_\text{fit}$ and therefore is primarily sensitive to point-like emission. Detecting extended structure (such as discs) would require a more sophisticated approach, yielding higher sensitivities.

\begin{table*}
\caption{Results of our kernel phase analysis when classifying the 1/3 most significant detections based on $\text{SNR}_\text{scaled}$ as candidate detections (``Can?'') and the rest as calibrators (``Cal?'') for the empirical detection method. Candidate detections with an empirical detection significance $\text{SNR}_\text{emp} > 5$ are classified as significant detections (``Det?''). K-band magnitudes are taken from SIMBAD.}
\begin{tabular}{l|c|c|c|c|c|c}
Name & $\text{SNR}_\text{ph}$ & $\text{SNR}_\text{scaled}$ & Can? & Cal? & $\text{SNR}_\text{emp}$ & Det? \\
\hline
IRAS 04108+2910 & 25.6 & 46.9 & N & Y & -- & N\\
FM Tau & 23.5 & 32.8 & N & Y & -- & N\\
CW Tau & 112.8 & 74.1 & N & Y & -- & N\\
FP Tau & 23.9 & 35.0 & N & Y & -- & N\\
CX Tau & 25.5 & 36.3 & N & Y & -- & N\\
2MASS J04154278+2909597 & 13.4 & 24.9 & N & Y & -- & N\\
CY Tau & 25.2 & 32.5 & N & Y & -- & N\\
V409 Tau & 21.3 & 33.6 & N & Y & -- & N\\
V410 Tau & 450.3 & 372.4 & Y & N & 122.7 & Y\\
BP Tau & 36.5 & 31.7 & N & Y & -- & N\\
V836 Tau & 23.6 & 30.4 & N & Y & -- & N\\
IRAS 04187+1927 & 131.8 & 130.5 & Y & N & 4.2 & N\\
DE Tau & 144.4 & 129.2 & Y & N & 2.0 & N\\
2MASS J04221675+2654570 & 25.0 & 39.1 & N & Y & -- & N\\
FT Tau & 33.3 & 43.0 & N & Y & -- & N\\
IP Tau & 26.4 & 30.4 & N & Y & -- & N\\
DG Tau & 1097.7 & 677.0 & Y & N & 3.5 & N\\
DH Tau & 28.3 & 30.2 & N & Y & -- & N\\
IQ Tau & 42.2 & 37.4 & N & Y & -- & N\\
DK Tau & 434.7 & 281.3 & Y & N & 1.8 & N\\
JH 56 & 10.5 & 14.8 & N & Y & -- & N\\
LkHa 358 & 52.9 & 112.8 & Y & N & 2.1 & N\\
HL Tau & 433.7 & 324.3 & Y & N & 4.9 & N\\
HK Tau & 38.1 & 49.1 & N & Y & -- & N\\
2MASS J04321540+2428597 & 85.9 & 88.2 & N & Y & -- & N\\
FY Tau & 66.8 & 67.2 & N & Y & -- & N\\
FZ Tau & 194.3 & 141.1 & Y & N & 2.6 & N\\
UZ Tau A & 318.8 & 231.9 & Y & N & 1.4 & \\
GI Tau & 43.2 & 40.2 & N & Y & -- & N\\
DL Tau & 66.9 & 64.5 & N & Y & -- & N\\
HN Tau A & 80.4 & 94.2 & N & Y & -- & N\\
DM Tau & 11.6 & 23.0 & N & Y & -- & N\\
CI Tau & 89.6 & 79.9 & N & Y & -- & N\\
IT Tau & 37.6 & 34.6 & N & Y & -- & N\\
AA Tau & 175.2 & 175.6 & Y & N & 2.3 & N\\
DN Tau & 46.5 & 45.9 & N & Y & -- & N\\
2MASS J04354093+2411087 & 80.1 & 94.9 & Y & N & 1.6 & N\\
HP Tau & 79.0 & 65.2 & N & Y & -- & N\\
DO Tau & 181.2 & 129.0 & Y & N & 3.2 & N\\
LkCa 15 & 117.2 & 124.0 & Y & N & 3.1 & N\\
JH 223 & 15.2 & 29.7 & N & Y & -- & N\\
GO Tau & 19.4 & 35.2 & N & Y & -- & N\\
DQ Tau AB & 58.9 & 57.3 & N & Y & -- & N\\
DR Tau & 272.8 & 159.3 & Y & N & 1.4 & N\\
DS Tau & 40.2 & 40.1 & N & Y & -- & N\\
GM Aur & 39.6 & 44.2 & N & Y & -- & N\\
MWC 480 & 1244.7 & 391.0 & Y & N & 1.6 & N\\
2MASS J05052286+2531312 & 13.0 & 54.8 & N & Y & -- & N\\
RW Aur A & 571.5 & 357.0 & Y & N & 3.6 & \\
V819 Tau & 17.1 & 20.4 & N & Y & -- & N\\
\end{tabular}
\label{tab:kernel_phase_analysis}
\end{table*}

\subsection{Comparison of Both Methods}

\begin{figure}
\centering
\includegraphics[width=\columnwidth]{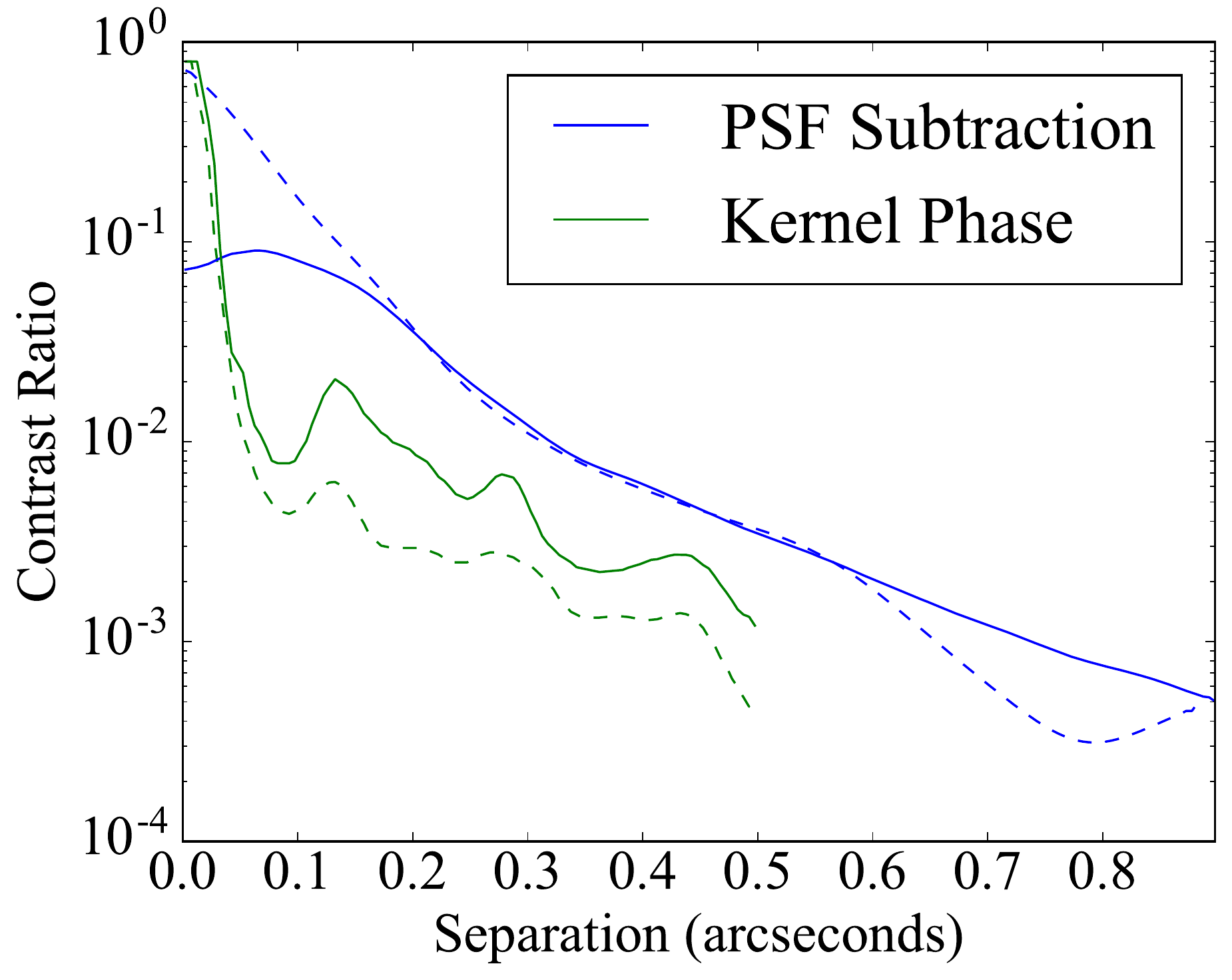}
\caption{The 5$\sigma$ contrast limit for AA Tau (\textbf{solid curve}) and CX Tau (\textbf{dashed curve)} as a function of separation from the star.  This shows, for example, at a separation of 0.3'' from AA Tau, the PSF subtraction method cannot detect anything less than 1/100 the brightness of the star but kernel phase can achieve limits of 1/300 the brightness of the star.}
\label{fig:bothContPlot}
\end{figure}

The 1D contrast plot (given for each separation by averaging around an annulus of fixed radius) is shown for both the PSF subtraction and kernel phase methods at the 5$\sigma$ level for 2 stars: AA Tau and CX Tau in Figure~\ref{fig:bothContPlot}. As theoretically expected, Figure~\ref{fig:bothContPlot} demonstrates that the kernel phase method outperforms the PSF subtraction method over its effective range of $\sim$0.5'' and the latter method is only useful at wider separations.  This was the case for all of our targets.  Our contrast limits for both methods indicate that we cannot detect objects fainter than $\sim$1/2000 the brightness of the star even at separations of 0.5'', which is insufficient for the detection of core-accreting giant planets \citep{wallace2019likelihood}.

\subsection{Significant features from kernel phase analysis}
\begin{figure*}
\centering
\includegraphics[width=0.49\textwidth]{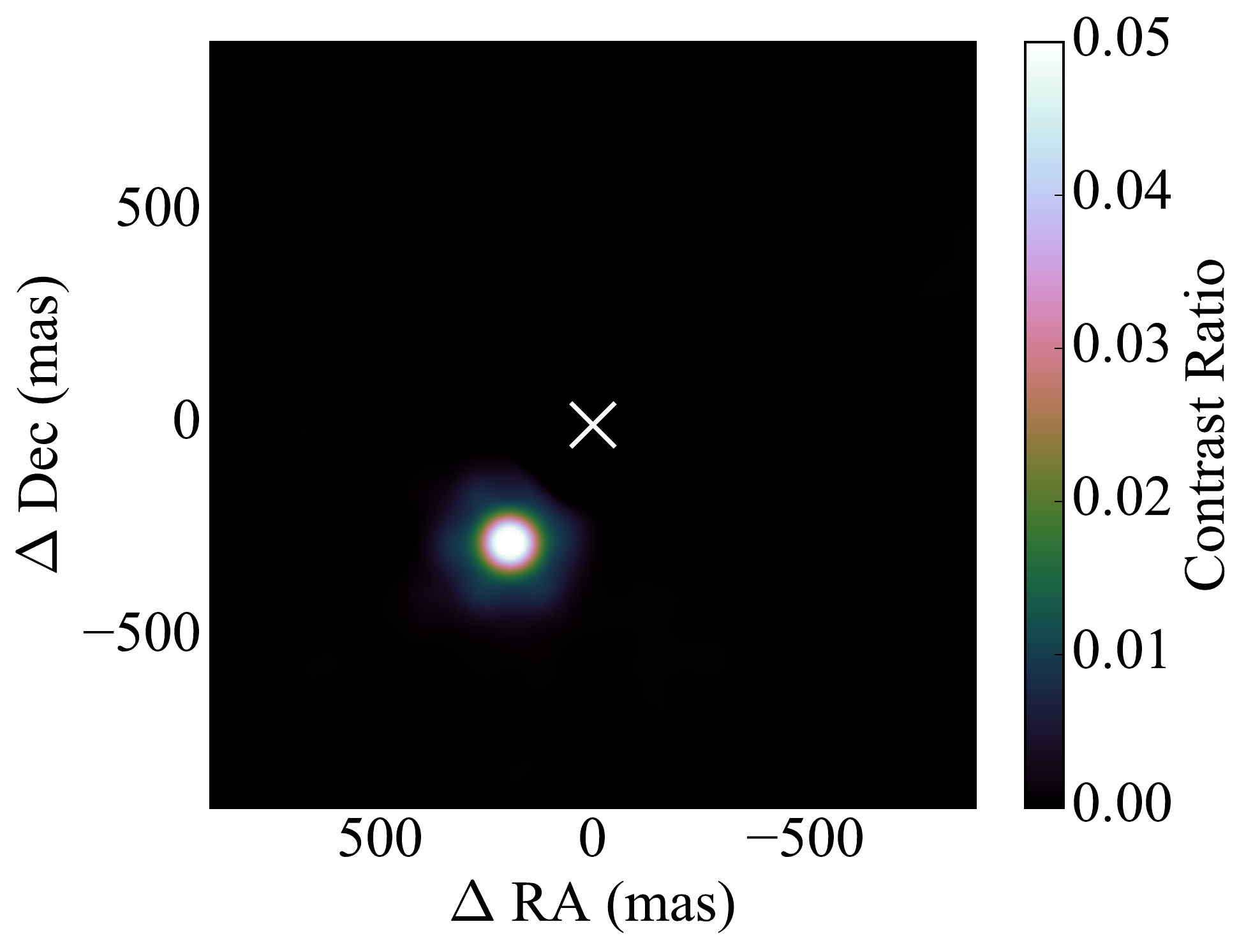}
\includegraphics[trim={10.25cm 0 10cm 0},clip,width=0.49\textwidth]{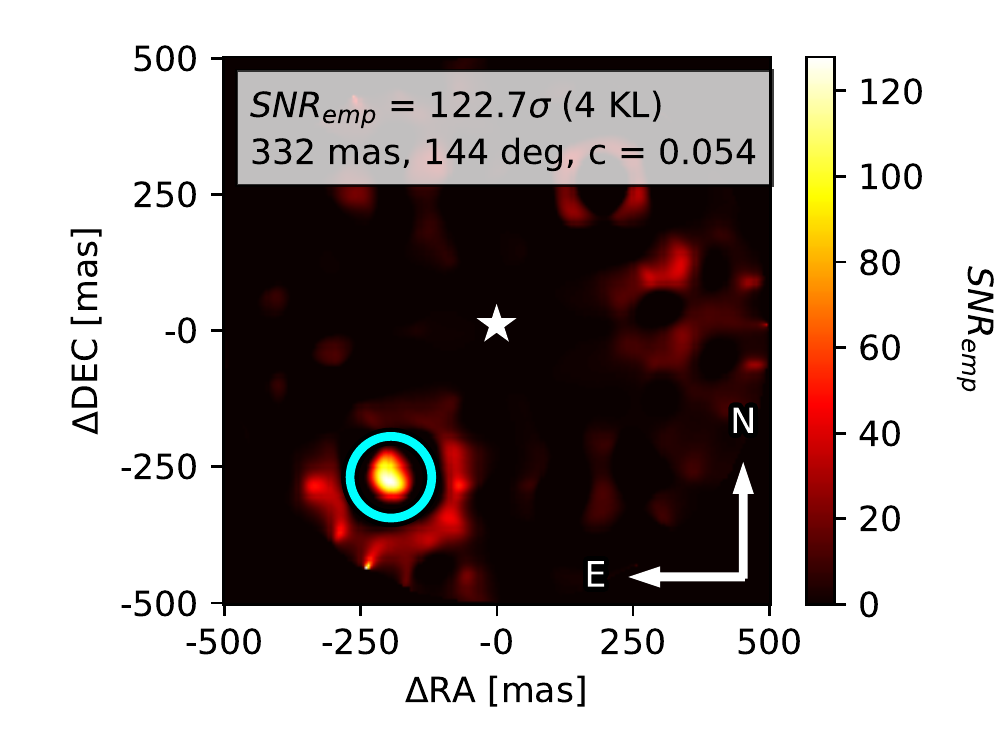}
\caption{Contrast map of V410 Tau using PSF subtraction (left panel) and kernel phase detection map (right panel). The host star is in the middle of the images and is removed by both methods. The companion is clearly visible to the south-east and consistently detected with both methods. In the kernel phase detection map, V410 Tau B's position is highlighted with a cyan circle and there is a residual halo around it which is caused by the limited Fourier coverage and model redundancies and disappears after subtracting the kernel phase signal of V410 Tau B from the data.}
\label{fig:V410TauMap}
\end{figure*}

\begin{figure*}
\centering
\includegraphics[width=10cm]{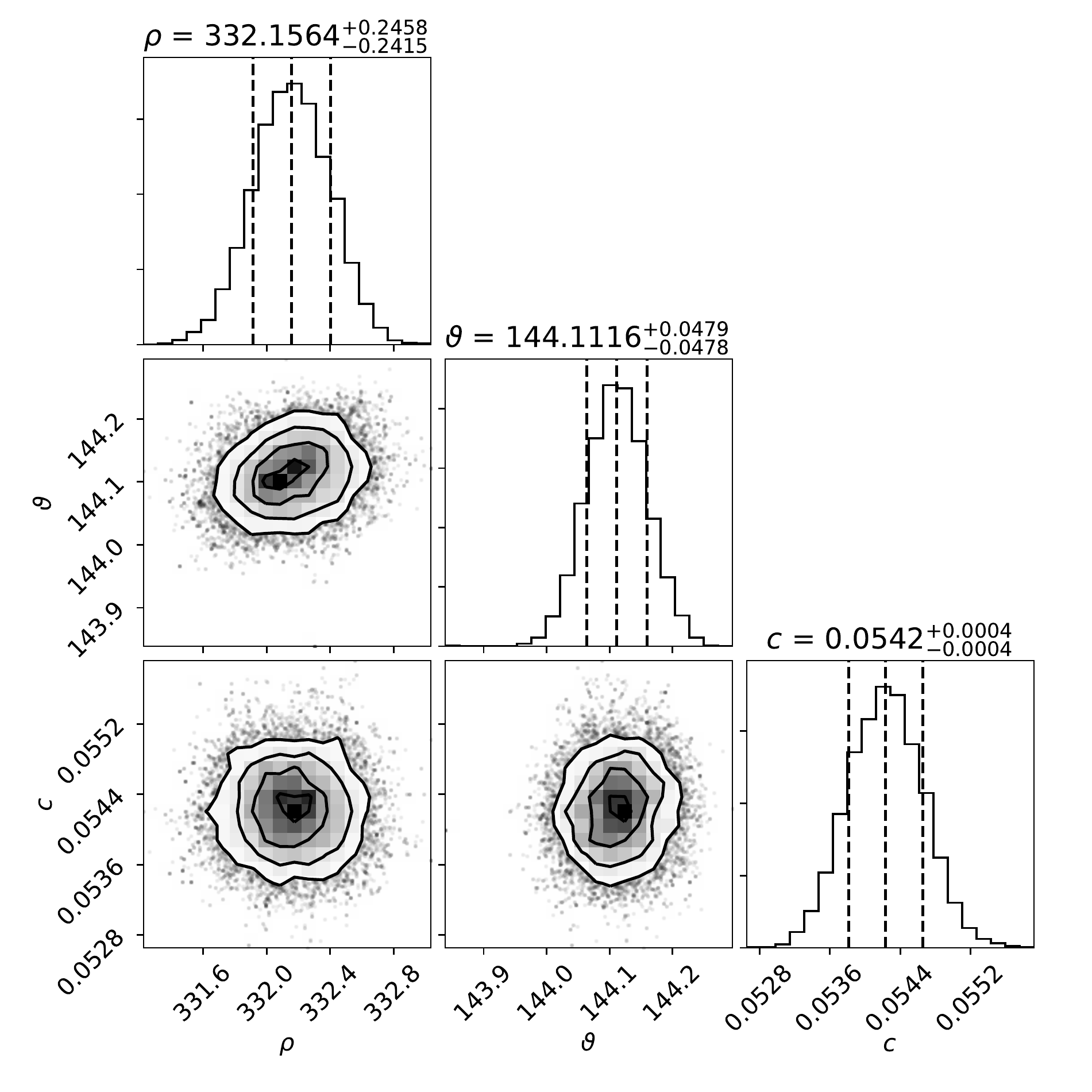}
\caption{Corner plot \citep{foreman-mackey2016} for an MCMC initialized around the best fit-position. The three fitted parameters are the angular separation $\rho$, the position angle $\theta$ and the contrast $c$. The MCMC is computed from the kernel phase using emcee \citep{foreman-mackey2013} with six random walkers initialized around the best-fit position and a temperature of $f_\text{err}^2 = (\sigma_\text{emp}/\sigma_\text{ph})^2 = 13.4 \approx \chi_\text{red}^2$, in order to find the best-fit parameters including their correlated uncertainties by maximizing the log-likelihood of the binary model \citep[cf.][]{kammerer2019kernel}.}
\label{fig:V410Corner}
\end{figure*}

The kernel phase analysis revealed several features, of which we define those with an empirical $\text{SNR}_\text{emp} > 5$ as significant detections. This criterion was only met by V410~Tau, which has a known brown dwarf companion \cite{ghez1997high}. We detect this known companion with both of our methods (cf., Figure~\ref{fig:V410TauMap}). The kernel-phase technique yields very precise constraints on its position and contrast, obtained from an MCMC fit (cf., Figure~\ref{fig:V410Corner}), and its best-fit parameters are listed in Table~\ref{tab:kerphase_companions}.

\begin{table}
\begin{tabular}{l|c|c|c}
Name & Sep. (mas) & Pos. Ang. ($^{\circ}$) & Contrast\\
\hline
V410 Tau B & $332.2\pm0.2$ & $144.11\pm0.05$ & $0.0542\pm0.0004$\\
\end{tabular}
\caption{Properties of the companion to V410 Tau with uncertainties from the kernel phase analysis.}
\label{tab:kerphase_companions}
\end{table}

Note that, as shown in Table~\ref{tab:kernel_phase_analysis}, HL Tau has a feature with $\text{SNR}_\text{emp} = 4.9$ so only just falls short of our detection threshold.  We believe this to be a feature of HL Tau's large protoplanetary disc \citep{partnership2015first} and not a companion.

% ==============================
\section{Wide Separation Analysis}
\label{sec:wide}
Due to a focus on efficient observations at small angles (see Section~\ref{sec:observations}), these data were mostly taken in a sub-array readout mode, limiting the field of view. We further extended our analysis to wider angles which were not covered by the 192 x 192 pixel cleaned images and analysis shown in Section~\ref{sec:psf_subtraction}. At these separations beyond $\sim$0.8", point spread function features were almost non-existent so we could use a more conventional image analysis with a simplified point-spread function model. 

To clean these full images, our image reduction simply consisted of dividing by a master flat and correcting bad pixels. The master flat was created from all dithered observations for a night, using pixels significantly away from detected objects. Companions were searched for over an $(\rho,\theta)$ grid in polar sky coordinates by performing aperture photometry with a simplified Gaussian PSF model, as described below. This truncated PSF model enabled searching for companions closer to the image edge. 

For each tested grid point, including the central star (i.e. separation $\rho$ of 0), we found a least squares solution to the flux $F$ of a model:

\begin{align}
    d_k &= B + F g_k,
\end{align}

where $d_k$ is the data for pixel $k$ over a 16 x 16 pixel grid, $g_k$ is a normalised Gaussian function with width matched to the observed PSF and $B$ the background. This least-squares flux solution for $F$ is simply given by:

\begin{align}
F &= \frac{\Sigma_k (g_k - \Sigma g_k/N) d_k}{\Sigma_k (g_k - \Sigma g_k/N) g_k},
\end{align}

where $N=256$, the total number of pixels. The uncertainty in pixels $k$ was simply estimated by the root mean square residuals of the fit, and the uncertainty in $F$ obtained by standard error propagation assuming independent background-limited uncertainties for all pixels. These fluxes were converted to contrasts by dividing by the fitted flux at a separation of 0, and these $(\rho,\theta)$ contrast maps averaged together with inverse variance weighting. Finally, uncertainties were corrected at each radius $\rho$ to ensure that the median absolute deviation of the residuals at every radius matched that of a unit Gaussian.

By highlighting features with significance greater than 7$\sigma$ and removing those that can be explained by the few residual speckles, we are able to determine the approximate positions of companions to our targets.  The contrast and position of each companion is calculated by fitting to the original reduced images.  We are able to detect significant companions for 9 of our objects, the properties of which are listed in Table~\ref{tab:companions}.

\begin{table*}
\caption{Properties of fitted companions}
\begin{tabular}{l|c|c|c|c}
Name & Separation ('') & Position Angle ($^{\circ}$) & Contrast ($\Delta m$) & \# Observations\\ \hline
2MASS J04354093+2411087 & 2.11 & 175.0 & 1.98$\pm$0.01 & 1\\
2MASS J05052286+2531312 & 2.35 & 59.8 & 2.24$\pm$0.01 & 2\\
DH Tau & 2.35 & 139.0 & 5.75$\pm$0.02 & 1\\
DK Tau & 2.39 & 119.5 & 1.81$\pm$0.01 & 4\\
HK Tau & 2.25 & 169.9 & 5.27$\pm$0.02 & 1\\
IRAS 04278+2253 & 1.29 & 95.9$\pm$0.6 & 2.00$\pm$0.02 & 1\\
IT Tau & 2.43 & 225.8 & 1.64$\pm$0.01 & 1\\
JH 223 & 2.15 & 342.2 & 2.40$\pm$0.01 & 3\\
RW Aur A & 1.49 & 254.6$\pm$0.1 & 2.42$\pm$0.01 & 2\\
\end{tabular}
\label{tab:companions}
\end{table*}

\begin{figure}
\centering
\subfigure[Fitted Contrast Map of DH Tau]{\label{fig:DHTauFit}\includegraphics[width=7cm]{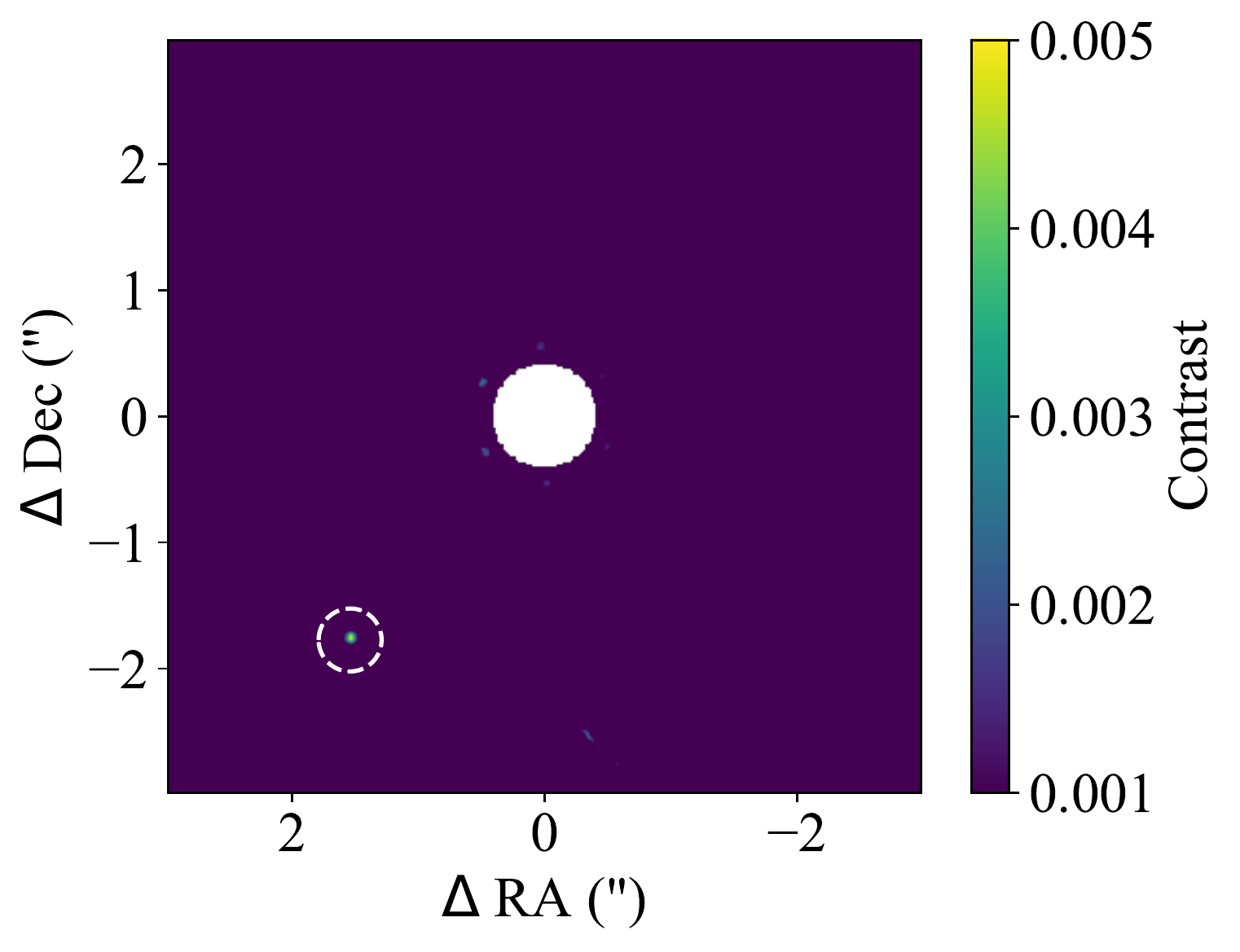}}
\subfigure[Fitted Contrast Map of HK Tau]{\label{fig:HKTauFit}\includegraphics[width=7cm]{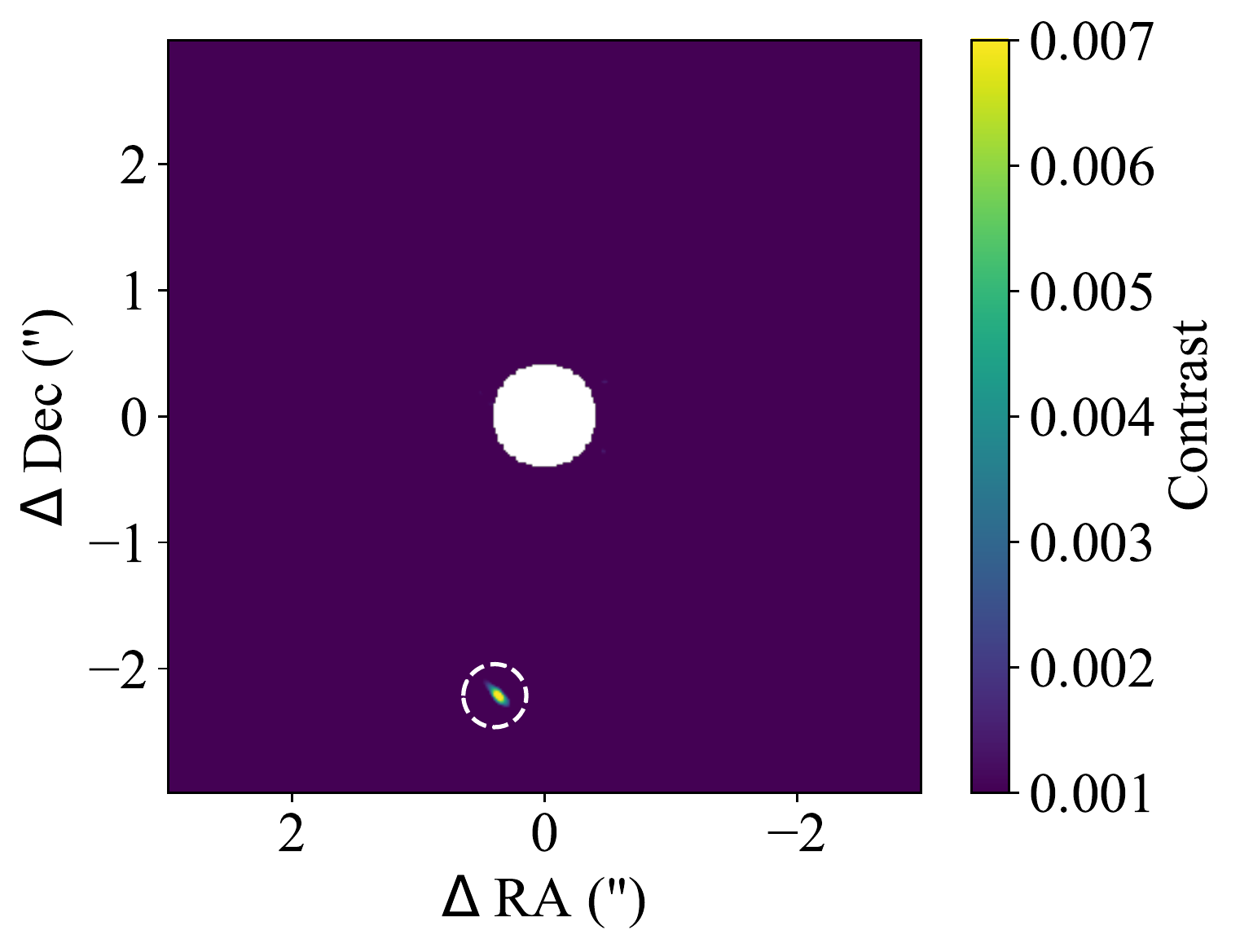}}
\caption{Contrast Maps of DH Tau and HK Tau with the companions circled.}
\label{fig:contFitDHHK}
\end{figure}

The companion of DH Tau is the only substellar-mass companion we are able to detect and our contrast is consistent with other studies such as \citet{kraus2013three}. We also find a high-contrast companion to HK Tau.  The fitted contrast maps for DH Tau and HK Tau are shown in Figure~\ref{fig:contFitDHHK}.

\begin{figure}
\centering
\includegraphics[width=\columnwidth]{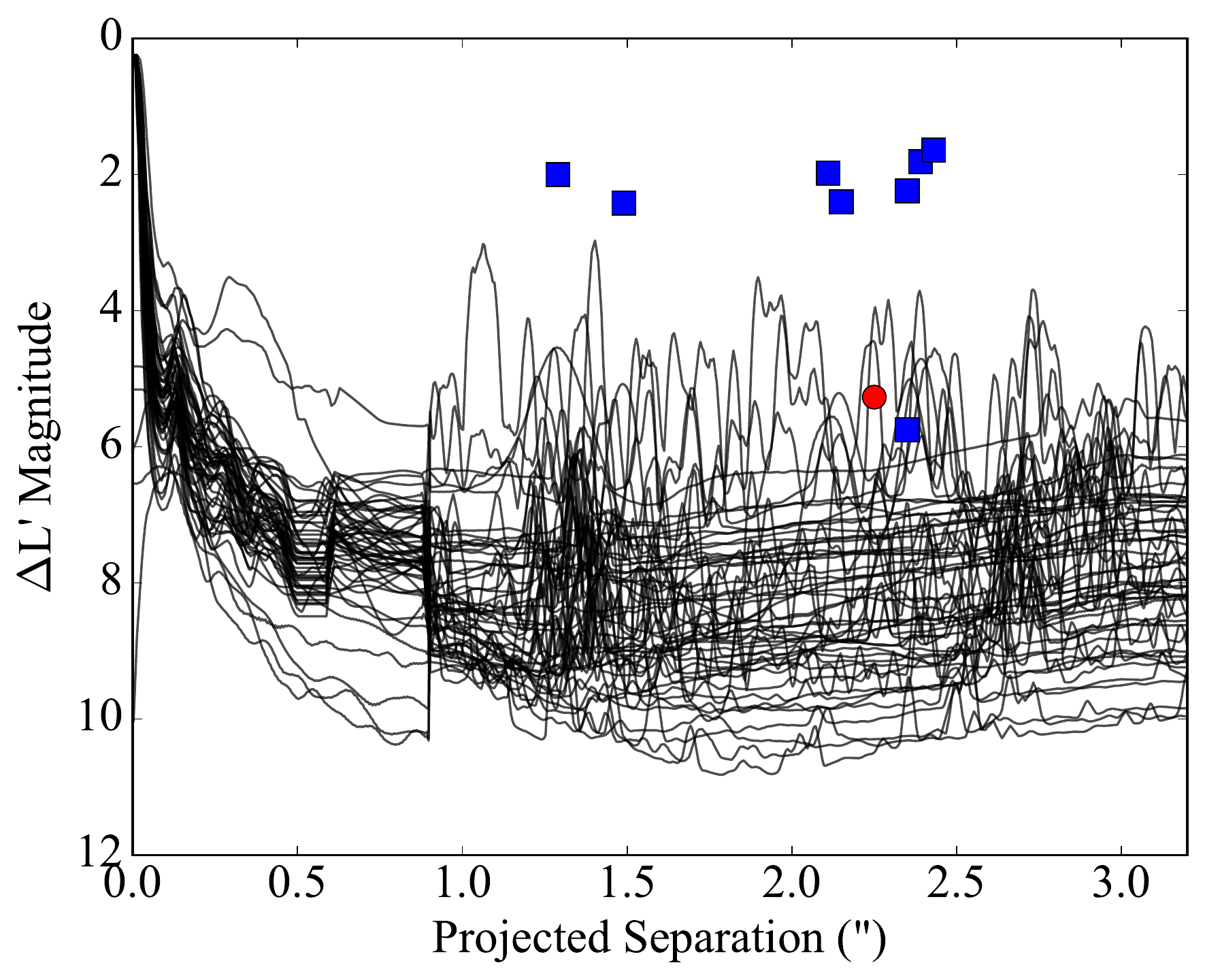}
\caption{5$\sigma$ contrast curves for all targets including contrast and separation of detected companions. Each line is our contrast limit for a particular target and the markers show detected companions.  The red circle indicates the companion to HK Tau.}
\label{fig:contAllPlot}
\end{figure}

Despite the high contrast of the companion to HK Tau, previous studies have concluded that it is a stellar-mass companion of similar spectral type to the primary and is obscured by an edge-on circumstellar disc \citep{stapelfeldt1998edge}. This circumstellar disc is represented by the elongated shape of the companion. Although we did not detect any additional brown dwarf mass companions to our targets, we now have a more complete picture of our contrast limits at wide separations. These contrast limits are listed in Table~\ref{tab:contrast}. Figure~\ref{fig:contAllPlot} shows the contrast curves for all targets with the detected companions indicated. The companion to HK~Tau is marked with a red circle and other companions are marked with blue squares.

\begin{table*}
\caption{Contrast limits for our targets using all three methods.}
\begin{tabular}{l|c|c|c|c|c|c|c}
& \multicolumn{7}{c}{Contrast Limit ($\Delta m$)} \\
Name & 0.1'' & 0.3'' & 0.5'' & 0.7'' & 1'' & 1.5'' & 2''\\\hline
IRAS 04108+2910 & 5.09 & 6.03 & 6.79 & 6.52 & 7.40 & 7.66 & 7.13\\
FM Tau & 5.49 & 6.64 & 7.10 & 6.85 & 7.87 & 7.85 & 7.67\\
CW Tau & 4.94 & 7.14 & 8.41 & 6.58 & 9.24 & 9.82 & 9.87\\
FP Tau & 5.54 & 7.06 & 8.04 & 7.26 & 7.93 & 7.65 & 7.57\\
CX Tau & 5.84 & 6.55 & 8.31 & 8.06 & 8.15 & 7.85 & 7.64\\
2MASS J04154278+2909597 & 5.68 & 6.02 & 7.01 & 6.40 & 6.35 & 6.79 & 6.35\\
CY Tau & 5.29 & 7.25 & 8.14 & 7.57 & 8.45 & 7.92 & 7.25\\
V409 Tau & 5.60 & 6.78 & 7.63 & 7.13 & 7.68 & 7.96 & 7.45\\
V410 Tau & 4.88 & 3.54 & 5.60 & 7.03 & 5.75 & 6.53 & 4.96\\
BP Tau & 5.54 & 7.51 & 8.49 & 7.89 & 9.43 & 9.00 & 8.82\\
V836 Tau & 5.00 & 6.59 & 7.43 & 7.12 & 8.05 & 7.98 & 7.86\\
IRAS 04187+1927 & 4.68 & 6.42 & 7.84 & 7.82 & 8.31 & 8.82 & 9.06\\
DE Tau & 4.67 & 6.70 & 7.96 & 7.59 & 9.02 & 9.13 & 8.98\\
RY Tau & 6.01 & 8.07 & 9.69 & 10.10 & 8.27 & 9.19 & 9.47\\
2MASS J04221675+2654570 & 5.59 & 6.39 & 7.61 & 7.50 & 8.39 & 8.76 & 8.20\\
FT Tau & 5.61 & 6.36 & 7.41 & 7.20 & 8.28 & 8.64 & 8.17\\
IP Tau & 6.29 & 7.09 & 8.16 & 6.90 & 7.51 & 7.27 & 7.37\\
DG Tau & 3.91 & 6.26 & 7.11 & 6.84 & 8.30 & 7.39 & 9.80\\
DH Tau & 6.05 & 6.15 & 7.93 & 7.62 & 8.57 & 8.06 & 8.47\\
IQ Tau & 6.57 & 6.49 & 8.08 & 7.83 & 8.89 & 8.94 & 8.77\\
UX Tau & 6.34 & 8.25 & 8.73 & 8.99 & 6.55 & 6.39 & 5.97\\
DK Tau & 5.05 & 6.24 & 7.71 & 7.72 & 8.69 & 9.11 & 9.13\\
IRAS 04278+2253 & 3.29 & 4.33 & 5.13 & 5.45 & 6.62 & 7.43 & 8.73\\
JH 56 & 6.56 & 7.15 & 7.64 & 6.80 & 7.81 & 7.73 & 7.42\\
LkHa 358 & 3.92 & 5.59 & 6.61 & 6.94 & 8.39 & 7.39 & 7.58\\
HL Tau & 4.90 & 6.16 & 7.37 & 7.26 & 8.47 & 10.36 & 10.43\\
HK Tau & 5.21 & 6.57 & 7.66 & 7.47 & 7.29 & 7.36 & 7.13\\
2MASS J04321540+2428597 & 5.10 & 6.59 & 7.56 & 7.53 & 8.75 & 9.60 & 9.38\\
FY Tau & 5.44 & 6.77 & 7.71 & 8.14 & 8.92 & 8.81 & 8.87\\
FZ Tau & 5.18 & 6.37 & 7.23 & 7.59 & 9.31 & 10.15 & 9.71\\
UZ Tau A & 3.99 & 5.89 & 7.01 & 7.08 & 9.21 & 9.62 & 9.01\\
GI Tau & 5.48 & 7.23 & 8.30 & 7.39 & 7.41 & 7.50 & 8.51\\
DL Tau & 5.20 & 7.12 & 8.18 & 8.04 & 9.29 & 9.51 & 9.21\\
HN Tau A & 5.07 & 6.47 & 7.37 & 7.79 & 8.70 & 9.07 & 8.92\\
DM Tau & 5.51 & 6.27 & 6.95 & 6.85 & 7.23 & 7.17 & 6.83\\
CI Tau & 5.71 & 7.24 & 8.05 & 8.40 & 9.02 & 9.36 & 9.17\\
IT Tau & 4.39 & 6.15 & 7.55 & 7.61 & 8.64 & 8.70 & 8.50\\
AA Tau & 4.99 & 6.01 & 7.25 & 7.33 & 8.51 & 8.42 & 8.16\\
DN Tau & 6.90 & 7.22 & 8.05 & 7.27 & 9.19 & 8.99 & 8.64\\
2MASS J04354093+2411087 & 5.21 & 6.13 & 6.87 & 7.63 & 8.91 & 9.07 & 7.51\\
HP Tau & 5.17 & 6.28 & 7.89 & 7.20 & 6.66 & 6.11 & 5.38\\
DO Tau & 4.86 & 6.05 & 7.88 & 7.34 & 8.62 & 9.36 & 9.70\\
LkCa 15 & 4.82 & 6.00 & 7.54 & 8.34 & 7.38 & 7.36 & 7.82\\
JH 223 & 6.11 & 6.92 & 7.36 & 7.16 & 7.44 & 7.10 & 6.71\\
GO Tau & 5.48 & 6.81 & 7.29 & 6.81 & 7.53 & 7.35 & 7.16\\
DQ Tau & 5.55 & 7.30 & 8.04 & 8.57 & 9.07 & 9.18 & 8.63\\
DR Tau & 4.82 & 6.32 & 7.57 & 7.96 & 9.14 & 10.30 & 10.09\\
DS Tau & 6.12 & 6.71 & 7.78 & 7.52 & 8.76 & 8.91 & 8.84\\
GM Aur & 5.27 & 6.99 & 8.25 & 7.62 & 8.37 & 8.07 & 8.09\\
AB Aur & 6.40 & 8.58 & 9.40 & 9.86 & 7.75 & 9.83 & 9.26\\
SU Aur & 5.72 & 7.95 & 9.25 & 9.75 & 8.73 & 8.16 & 8.09\\
MWC 480 & 4.76 & 6.13 & 6.81 & 7.34 & 8.86 & 10.34 & 10.52\\
2MASS J05052286+2531312 & 5.00 & 6.04 & 6.53 & 6.59 & 6.63 & 6.49 & 6.84\\
RW Aur A & 4.55 & 5.92 & 7.02 & 7.10 & 9.06 & 5.23 & 10.32\\
V819 Tau & 6.07 & 7.08 & 7.92 & 7.63 & 8.42 & 8.36 & 7.97\\
\end{tabular}
\label{tab:contrast}
\end{table*}

Assuming an age of 1\,Myr for our planets, which is conservative as they may still be forming in a Class II disc, we converted the contrast into a mass limit using models from \citet{spiegel2012spectral}.  For this conversion we need to assume an appropriate internal entropy for our planets. As mentioned previously, planet luminosity and internal entropy is highly uncertain but recent models suggest few Jupiter mass planets have initial entropy no less than $\sim$10--11\,$k_\mathrm{B}$/baryon \citep[e.g.][]{mordasini2013luminosity,berardo2017evolution,Marleau19}. The hot-start and cold-start entropy curves take the form of a `tuning fork' with hot-start entropy increasing with mass and cold-start entropy decreasing with mass \citep{marley2007luminosity}. However, since hot-start models are expected to be more likely for high-mass planets due to the difficulty in radiating away the accretion luminosity for all but the lowest accretion rates, a reasonable assumption is that the average entropy is fairly constant at somewhere around 10--11\,$k_\mathrm{B}$/baryon across the 1--10\,M$_J$ range. To keep our model simple, we assume a single value of initial internal entropy, regardless of mass. Analysis of directly imaged planets indicates $\beta$-Pic b, at the high-mass end of the planet distribution, formed with a minimum entropy of $\sim$10.5\,$k_\mathrm{B}$/baryon \citep{marleau2014constraining}. We have decided to use this value to calculate mass limits as it is also close to the average initial entropy of a 1\,M$_{\rm{J}}$ planet according to \citet{spiegel2012spectral} and can be applied to a wide temperature range of 500--1500\,K \citep{berardo2017evolution}.  Using this entropy, we calculate planet magnitude as a function of mass and age using \citet{spiegel2012spectral}, and thus convert contrast ratio to mass.

\begin{figure}
\centering
\includegraphics[width=\columnwidth]{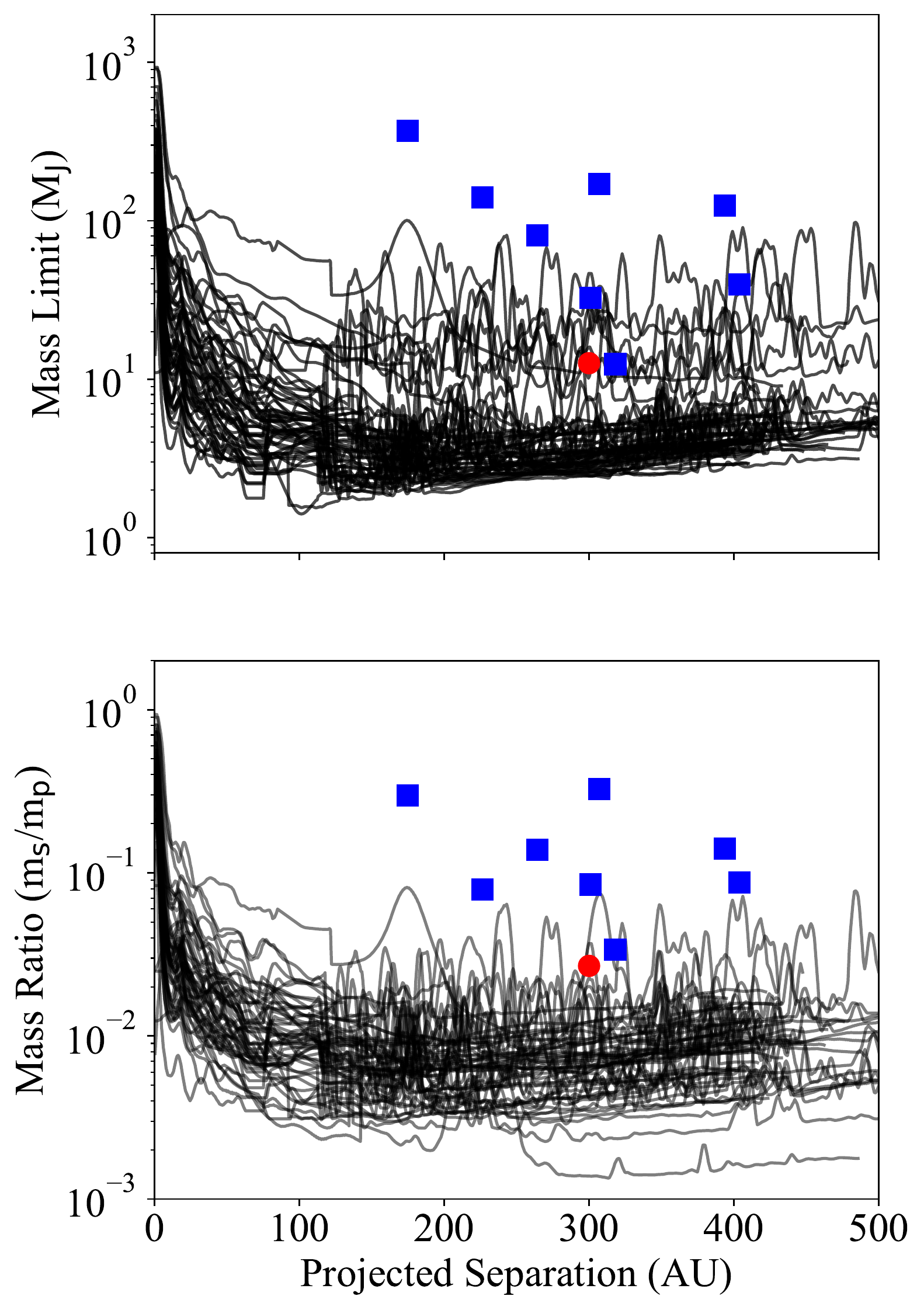}
\caption{Mass and mass ratio limits for all targets.  The companion to HK Tau is again represented by a red circle as the actual mass is assumed to be higher than shown here.}
\label{fig:massAllPlot}
\end{figure}

Using the stellar masses shown in Table~\ref{tab:allObs}, this mass limit was also converted to a mass ratio.  This is shown for all targets in Figure~\ref{fig:massAllPlot}.
As shown in Figures~\ref{fig:contAllPlot} and~\ref{fig:massAllPlot}, the companion to DH Tau is close to the faintest we were able to detect.  Note while the companion to HK Tau is included (shown with a red circle,) the circumstellar disc reduces its brightness so the true mass is probably much higher than that shown in Figure~\ref{fig:massAllPlot}.  
The top panel of Figure~\ref{fig:massAllPlot} shows that we are able to detect planetary-mass companions (<13\,M$_{\rm{J}}$) for most of our targets at wide separations (>100\,au). The lack of new brown dwarf detections from our data implies these companions are rare at wide separations, providing evidence of the ``brown dwarf desert'' described by \citet{marcy2000planets} and \citet{grether2006dry}. The lack of planetary-mass detections allows us to constrain the maximum frequency of hot-start planets in the TMC.

% ========================================
\section{The Frequency of Wide Separation Massive Planets}
\label{sec:freq}

\subsection{Total Probability of Planet detection}
\begin{figure}
\centering
\includegraphics[width=\columnwidth]{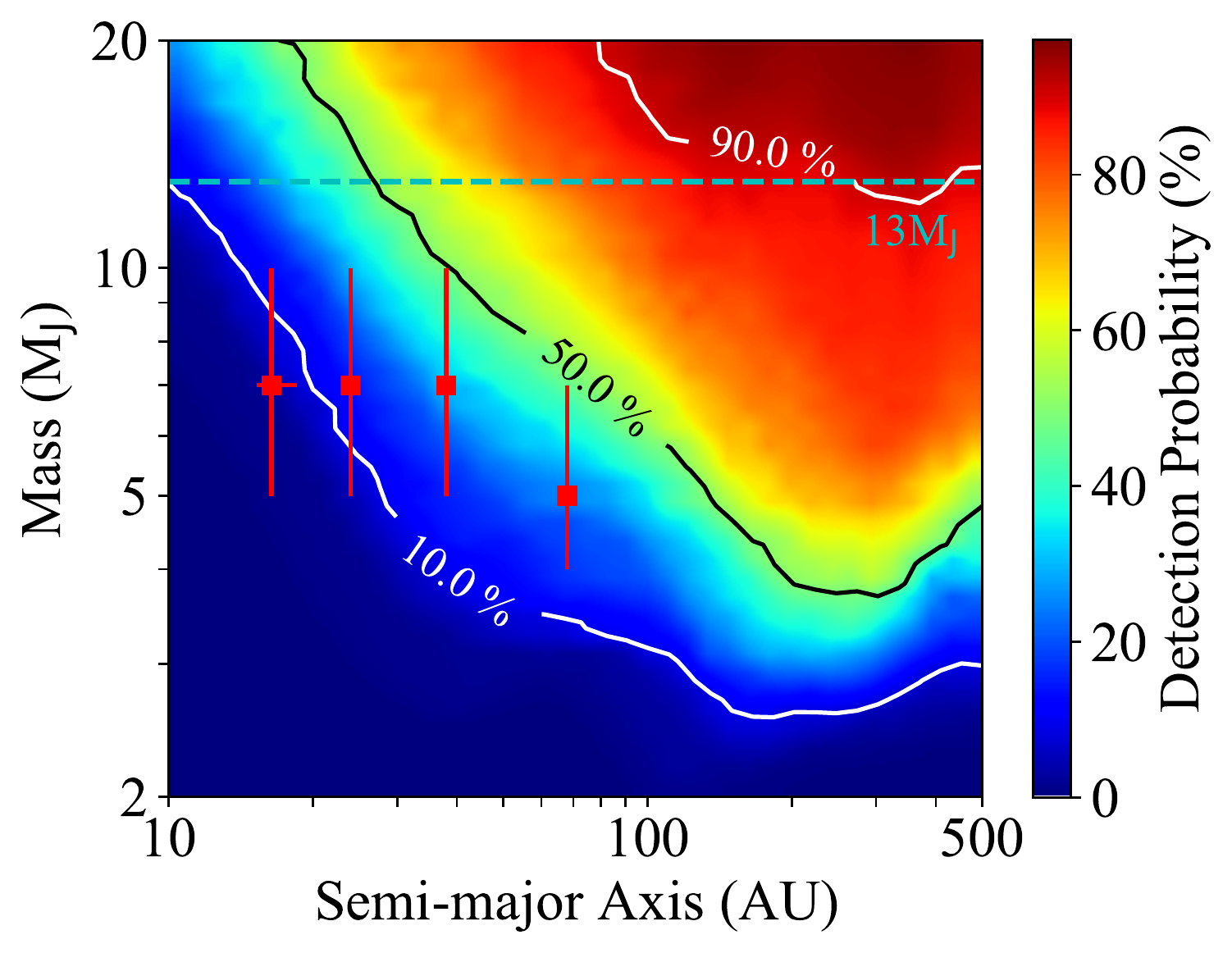}
\caption{Probability of planet detection as a function of mass and semi-major axis for a planet age of 1\,Myr and initial internal entropy of 10.5\,$k_\mathrm{B}$/baryon. The HR~8799 planets and the planet-mass threshold of 13\,M$_{\rm{J}}$ are also shown.}
\label{fig:probPlotAll}
\end{figure}

Despite our lack of planet detections, Figure~\ref{fig:massAllPlot} shows that our limits are sufficient for the detection of young planetary-mass companions for many of our targets. This opens up the possibility of detecting wide systems analogous to HR~8799. Combining the contrast limits for all of our targets, we can determine the likelihood of detecting a planet as a function of mass and semi-major axis. We apply the same method as Figure~\ref{fig:massAllPlot} with an age of 1\,Myr and initial entropy of 10.5\,$k_\mathrm{B}$/baryon to convert magnitude to mass using models from \citet{spiegel2012spectral}. using Monte-Carlo sampling, we randomise the system inclination and planet positions to get a more comprehensive view of our capabilities. This is shown in Figure~\ref{fig:probPlotAll}. The HR~8799 planets are shown, as well as the 13\,M$_{\rm{J}}$ planet-mass threshold.

\subsection{Comparison with HR 8799 analogues}
\begin{figure}
\centering
\includegraphics[width=\columnwidth]{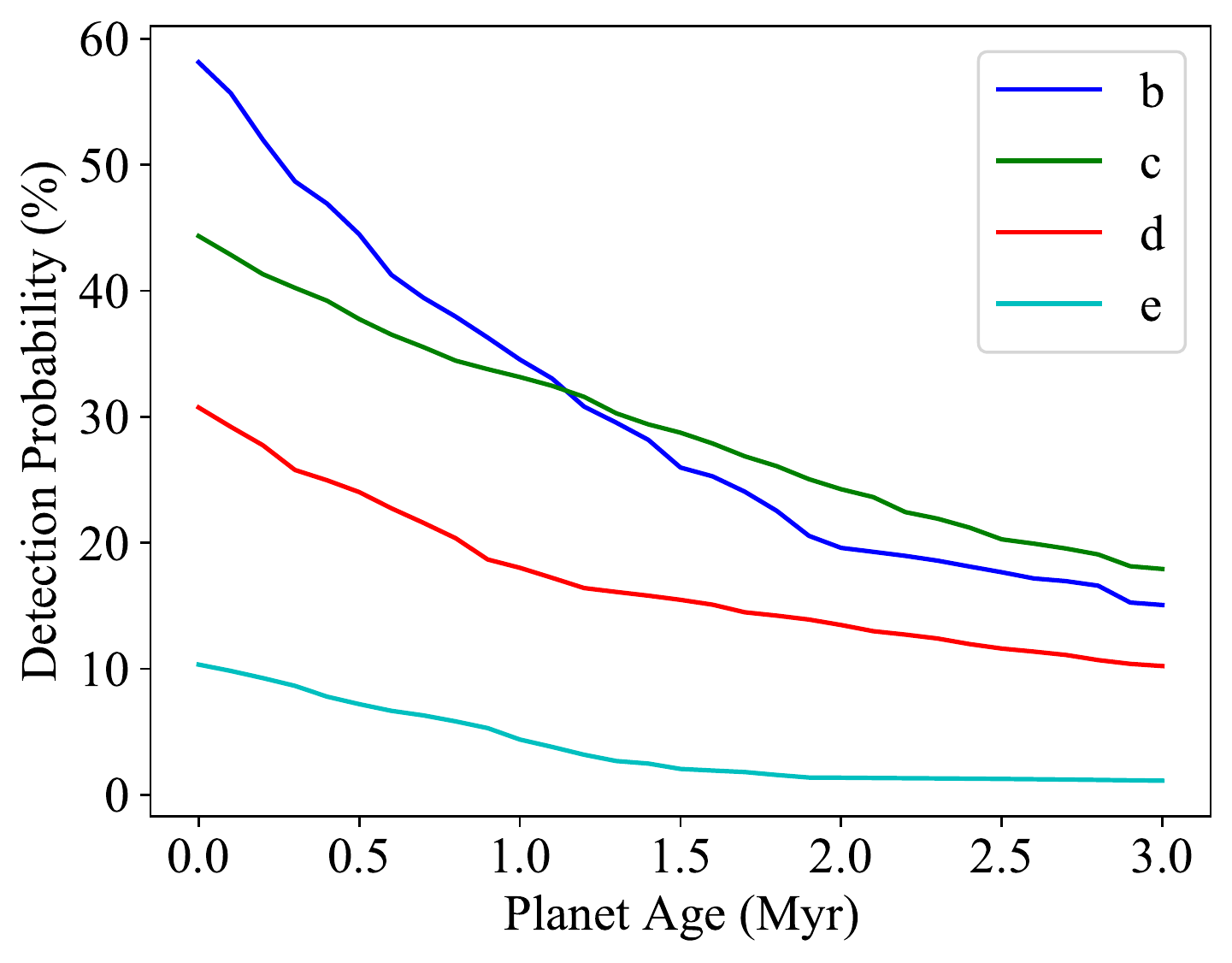}
\caption{Detection probability of HR~8799 analogues (averaged over all targets) versus age of the planets.  An age of 0 corresponds to the moment the planets stop accreting.}
\label{fig:hr8799}
\end{figure}

The result in Figure~\ref{fig:probPlotAll} shows that, averaged over all targets, we have a greater than 80\% probability of detecting >10\,M${\rm{J}}$ planets at separations beyond 100\,au. However, even at the lower mass and separations of an HR~8799 analogue, we still have a $\sim$20\% chance of detecting this system at an age of 1\,Myr with an initial internal entropy of 10.5\,$k_\mathrm{B}$/baryon. Applying the luminosity curves from \citet{spiegel2012spectral} to our HR~8799 analogue, we determine the probability of detecting these planets at ages of 0--3\,Myr.  Our detection probability of the 4~planets around HR~8799 is shown in Figure~\ref{fig:hr8799}.  

The curves in Figure~\ref{fig:hr8799} demonstrate how the planets around HR~8799 cool and fade over time. When the planets are newly formed, we have a greater than 40\% chance of detecting HR~8799~b and~c~analogues. At an age of 3\,Myr, we only have a 30\% chance of detecting these planets.

The stars in our sample are believed to have an age of $\sim$2--3\,Myr, which implies any planets around our targets are not much older than $\sim$1\,Myr. Since no planets were detected in our sample, we can make a statement on the maximum frequency of wide and massive systems.

\subsection{Planet frequency}
\begin{figure}
\centering
\includegraphics[width=\columnwidth]{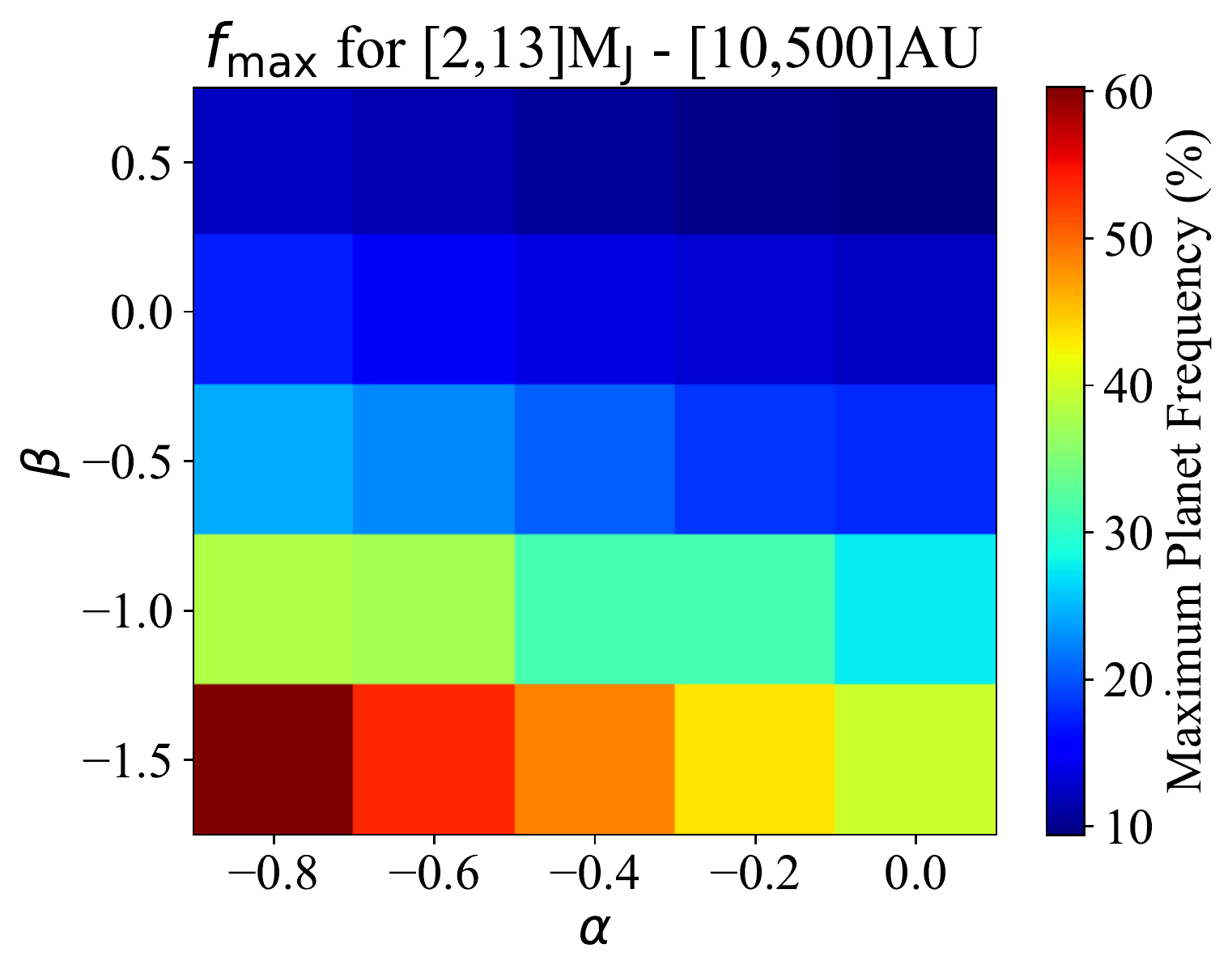}
\caption{Maximum planet frequency for mass 2--13\,M$_{\rm{J}}$ and semi-major axis 10--500\,au at 90\% confidence with differing mass and semi-major axis power-law indices ($\alpha$ and $\beta$ respectively.)}
\label{fig:frequency_indices}
\end{figure}

We use the same method from \citet{vigan2012international} to calculate the maximum planet frequency in a given range. This method assumes the likelihood of the data $d$ for a given frequency $f$ is given by
\begin{equation}
L(\left\lbrace d_{i}\right\rbrace |f) = \prod\limits_{i}^{N}(1-fp_{i})^{1-d_{i}}(fp_{i})^{d_{i}},
\end{equation}
where $N$ is the number of targets, in our case~55, $d_{i}$ is 0 if no planets are detected and 1 if at least 1~planet is detected. The probability $p_{i}$ is the probability of detecting planets in a given range assuming an appropriate planet distribution. We assume a power-law distribution in mass and semi-major axis such that
\begin{equation}
    \frac{dN_{\rm{planets}}}{d\mathrm{ln} Md\mathrm{ln}a} = CM^{\alpha}a^{\beta}.
\end{equation}
We obtain the posterior distribution from Bayes' theorem:
\begin{equation}
p(f|\left\lbrace d_{i}\right\rbrace) = \frac{L(\left\lbrace d_{i}\right\rbrace |f)p(f)}{\int\limits_{0}^{1}L(\left\lbrace d_{i}\right\rbrace |f)p(f)df},
\end{equation}
where $p(f)$ is the priori probability density of the frequency $f$ which we set to a uniform value of 1.  For a given confidence level, the maximum frequency is obtained using
\begin{equation}
\mathrm{Confidence} = \int\limits_{f_{\text{min}}}^{f_{\rm{max}}}p(f|\left\lbrace d_{i}\right\rbrace)df.
\end{equation}
We set $f_{\rm{min}}$ to 0 and rearrange to find $f_{\rm{max}}$.  This value was calculated over a semi-major axis range of 10--500\,au, the same as Figure~\ref{fig:probPlotAll}, and a mass range of 2--13\,M$_{\rm{J}}$.  Since we did not detect any planets in this range, all of our values of $d_{i}$ will be 0. To obtain our probabilities $p_{i}$, we try several values for the mass and semi-major axis power-law indices while keeping the range constant at [2,13]\,M$_{\rm{J}}$ and [10,500]\,au. The planet age is again set to be 1\,Myr and the initial internal entropy is 10.5\,$k_\mathrm{B}$/baryon. The maximum frequency at a 90\% confidence level is shown in Figure~\ref{fig:frequency_indices}.

As shown in Figure~\ref{fig:frequency_indices}, the maximum frequency is better constrained at higher power-law indices but these are considered unlikely power-law indices at this mass and semi-major axis range. The symmetric power law given by \citet{fernandes2019hints} has a mass index of $\alpha=-0.45$ and a semi-major axis index of $\beta=-0.95$. The study from \citet{bowler2018occurrence} has $\alpha=-0.65$ and $\beta=-0.85$. Our result shows that less than $\sim$30\% of stars have a planet in this mass and semi-major axis range if we assume one of these power-law distributions.

\begin{figure}
\centering
\includegraphics[width=\columnwidth]{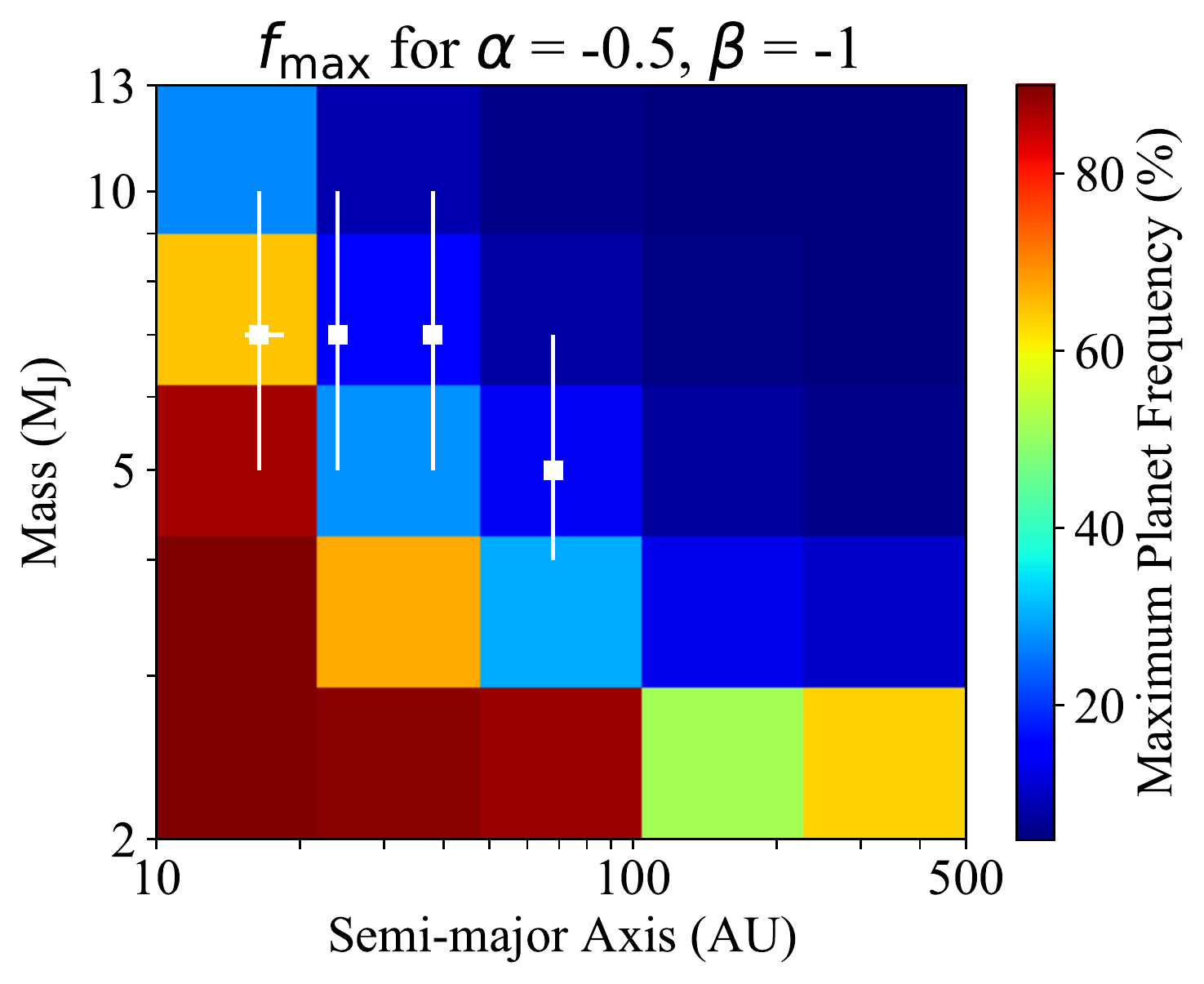}
\caption{Maximum planet frequency assuming a power law with $\alpha=-0.5$ and $\beta=-1$ at 90\% confidence. The white points show the planets around HR~8799.}
\label{fig:frequency_mass_sep}
\end{figure}

Assuming a power-law distribution in which $\alpha=-0.5$ and $\beta=-1$, we also calculate the dependence of planet frequency on mass and semi-major axis. This is shown in Figure~\ref{fig:frequency_mass_sep} over a mass range of 2--13\,M$_{\rm{J}}$ and semi-major axis range of 10--500\,au at a 90\% confidence level. The planets around~HR 8799 are also marked.

The result from Figure~\ref{fig:frequency_mass_sep} confirms that massive planets at wide separations are very rare, occurring around less than 10\% of stars. Planets with the mass and semi-major axis similar to HR~8799~b, c~and~d are expected to occur around less than 20\% of stars, while analogues to HR~8799~e may be more common, but we cannot draw a strong conclusion from our results regarding this aspect.

%================================
\section{Summary and Conclusions}
\label{sec:conclusion}
In this work, we have conducted a high-contrast imaging survey of the Taurus molecular cloud with the aim of finding any massive young planets and planets in the process of forming. Using the PSF subtraction technique, we found that our limits are not sufficient to detect planetary-mass companions at small separations. The kernel phase method improved our limits at small separations, but was still insufficient for detection of solar-system analogues. For non-accreting planets, our detection limits were similar to \citet{kraus2011mapping} at 20\,au ($\sim$15\,M$_J$ median mass limit), but a factor of 10 deeper in mass at 150\,au ($\sim$3\,M$_J$ median mass limit).  Our probabilities of planet detection as a function of mass and semi-major axis are broadly comparable to the result from SHINE, the SPHERE infrared survey \citep{vigan2020sphere} which used a larger sample of targets.

The continued lack of new brown dwarf companions at wide separations is further evidence of the so-called ``brown dwarf desert'' described by \citet{marcy2000planets} and \citet{grether2006dry} extending to separations beyond that probed by radial velocity surveys.  We were able to detect several known wide companions, including the roughly planetary-mass companion DH~Tau~b and the circumstellar disc around the companion to HK Tau.

We determined that, if the HR~8799 planets were placed in the TMC at the appropriate age, we could have detected analogues to HR~8799~b, c,~and~d around more than 15\% of our targets at an age of 1\,Myr. Assuming a similar power law to \citet{fernandes2019hints}, we find that planets with the mass or semi-major axis of HR~8799~b, c,~and~d occur around less than 20\% of stars.  Generalising this to planets from 2--13\,M$_{\rm{J}}$ at separations 10--500\,au, we found that, assuming the same power law, the planet frequency in this mass and semi-major axis range is less than 30\% at a 90\% confidence level. Future instruments such as VIKiNG on VLTI and METIS on the E-ELT will be required to improve on our detection limits, to more precisely constrain planet frequency.

\section*{Data Availability}
The data underlying this article are available from the corresponding author on reasonable request.

\section*{Acknowledgements}
We thank the anonymous referee for their useful comments which greatly improved this study.

The data presented herein were obtained at the W.~M.~Keck Observatory, which is operated as a scientific partnership among the California Institute of Technology, the University of California and the National Aeronautics and Space Administration.  The observatory was made possible by the generous financial support of the W.~M.~Keck Foundation.

A.L.W. would like to thank the Australian Government for their support through the Australian Postgraduate Award Scholarship and the Research School of Astronomy and Astrophysics at the Australian National University for the ANU Supplementary PhD Scholarship.

M.J.I. acknowledges funding provided by the Australian Research Council (Discovery Project DP170102233 and Future Fellowship FT130100235)

C.~F.~acknowledges funding provided by the Australian Research Council (Discovery Project DP170100603 and Future Fellowship FT180100495), and the Australia-Germany Joint Research Cooperation Scheme (UA-DAAD).

\renewcommand\refname{References}
\bibliographystyle{mnras}
\bibliography{references}

%%%%%%%%%%%%%%%%%%%%%%%%%%%%%%%%%%%%%%%%%%%%%%%%%%

%%%%%%%%%%%%%%%%% APPENDICES %%%%%%%%%%%%%%%%%%%%%

%%%%%%%%%%%%%%%%%%%%%%%%%%%%%%%%%%%%%%%%%%%%%%%%%%

% Don't change these lines
\bsp	% typesetting comment
\label{lastpage}
\end{document}